# MULTIPLE SPACE DEBRIS COLLECTING MISSION

# DEBRIS SELECTION AND TRAJECTORY OPTIMIZATION

Max CERF

EADS Astrium Space Transportation

78130 Les Mureaux, France

max.cerf@astrium.eads.net

## ABSTRACT

A possible mean to stabilize the LEO debris population is to remove each year 5 heavy debris like spent satellites or launchers stages from that space region. This paper investigates the $\Delta V$ requirement for such a Space Debris Collecting mission. The optimization problem is intrinsically hard since it mixes combinatorial optimization to select the debris among a list of candidates and functional optimization to define the orbital maneuvers. The solving methodology proceeds in two steps : firstly a generic transfer strategy with impulsive maneuvers is defined so that the problem becomes of finite dimension, secondly the problem is linearized around an initial reference solution. A Branch and Bound algorithm is then applied to optimize simultaneously the debris selection and the orbital maneuvers, yielding a new reference solution. The process is iterated until the solution stabilizes on the optimal path. The trajectory controls and dates are finally re-optimized in order to refine the solution. The method is applicable whatever the numbers of debris (candidate and to deorbit) and whatever the mission duration. It is exemplified on an application case consisting in selecting 5 SSO debris among a list of 11.

I  **INTRODUCTION**

   **A) Space debris**

A drastic growth of the space debris population in the LEO region under 2000 km is foreseen in the next decades with high collision risks for future space flights[13,15]. A particularly crowded region is the vicinity of the SSO and polar orbits in the range 600-1200 km altitude and 80-105deg inclinations with a density peak around the altitudes of 800-900 km. These orbits are well adapted to Earth observation and are therefore intensively used. The most efficient ways to mitigate this scenario are to avoid the creation of new debris by appropriate spacecrafts conception and to actively remove the existing debris[1,18]. Several studies have led to the conclusion that removing 5 heavy LEO debris per year, like spent satellites or launchers upper stages, is mandatory to stabilize the debris population[2]. This paper deals with such a Space Debris Collecting mission that would meet this requirement.

   **B) Mission definition**

The Space Debris Collecting (SDC) mission aims at deorbiting 5 heavy debris per year. The debris must be selected in a list of N candidates, so that the required propellant consumption for the mission is minimized.

The deorbitation consists in clearing the LEO region (altitude below 2000 km) : this can be achieved by either making the debris re-enter the atmosphere (preferred solution if possible) or re-orbiting the debris at a higher altitude[1]. For the SSO debris considered in this paper, only the reentry solution is envisioned.

Two deorbitation scenarios can be envisioned after the vehicle has captured the debris :

- Either the vehicle realizes the deorbitation manoeuvre. The debris is then released on the reentry trajectory while the vehicle returns on a stable orbit in order to pursue the mission.
- Or the vehicle supplies the debris with a deorbitation device (booster, tether, …). The debris performs then the deorbitation manoeuvre by itself, while the vehicle pursues the mission from the initial debris orbit.

The mission starts from the first debris selected on the path. It is assumed that the launcher has realized the rendezvous maneuvers in order to bring the SDC vehicle to this first debris. The cost and duration to reach this first debris are therefore not counted into the mission cost and duration.

### C) Global optimization problem

The global SDC problem is a mix of hard embedded optimization problems :

- several continuous transfer problems consisting in optimizing the trajectory from one debris to the other.
- a combinatorial path problem consisting in selecting the debris and the collecting order.

Table 1 gives an overview of these problems features.

|  | Transfer problem | Path problem |
|---|---|---|
| Goal | Optimize the trajectories | Select the debris and the order |
| Category | Optimal control theory | Graphs theory |
| Algorithms | Non linear programming | Integer programming |
| Complexity | Functional optimization<br>→ infinite dimension<br>→ hard continuous problem | NP problem (TSP like)<br>→ exponential<br>→ hard combinatorial problem |
| Problem nature | **Non linear time dependent graph problem** | |

**Table 1 : Two embedded optimization problems**

It will not be possible to tackle directly the global problem in its general complexity. The methodology to solve the problem follows the successive steps :

- Write the global problem formulation (Part II)
- Investigate the existing numerical methods in order to choose the most adapted
- Simplify the transfer problem considering the mission specificities (Part III)
- Linearize the formulation and iterate on the local solution (Part IV)
- Issue the optimal path (debris selection and order)
- Reoptimize the controls and dates on the selected path

This solving methodology is depicted on Figure 1.

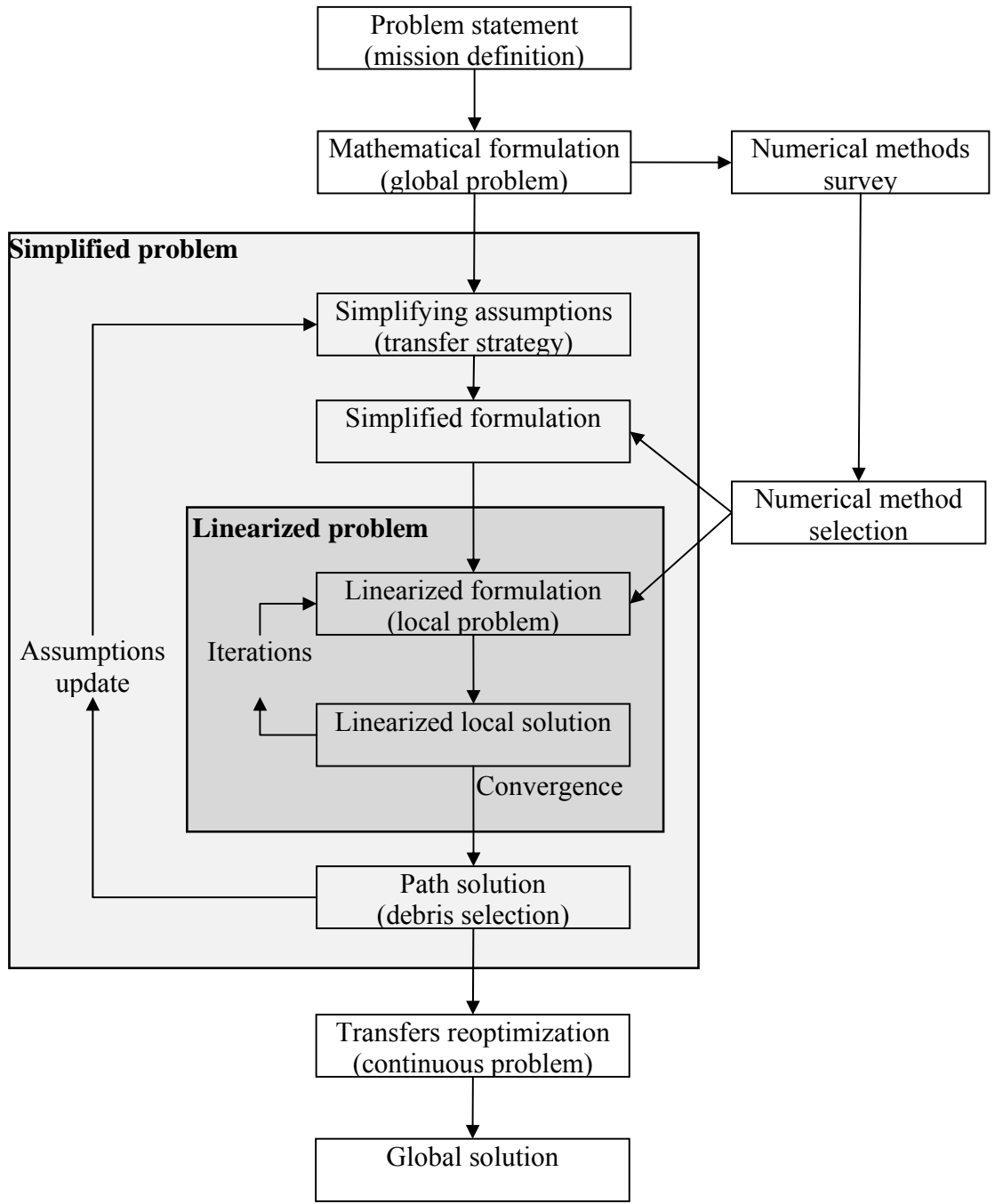

**Figure 1 : Solving methodology**

## II   PROBLEM FORMULATION

This section presents the mathematical formulation of the two embedded optimization problems and gives an overview of the available numerical methods.

### A) Transfer problem

The transfer problem consists in going from a debris i to another debris j. In terms of optimal control theory, the SDC vehicle is represented by a 7-state vector X(t). The components of this state vector are :

$$X(t) = \begin{pmatrix} \vec{r}(t) = \text{position} \\ \vec{v}(t) = \text{velocity} \\ m(t) = \text{mass} \end{pmatrix} = \begin{pmatrix} Y(t) \\ m(t) \end{pmatrix} \qquad (1)$$

To ease the subsequent formulation of the optimal control problem, we introduce the 6-vector Y(t) representing the 6 first components of X(t), i.e. the position and the velocity. An equivalent state representation is given by the 6 orbital parameters instead of the 3 position and the 3 velocity components. This equivalent representation will be considered when simplifying the problem formulation.

The 6-state vector of the debris k at the date t defining the debris position and velocity (or orbital parameters) is similarly denoted $Y_k(t)$.

The SDC vehicle trajectory is controlled by the 3-command vector U(t) representing the vehicle thrust at each instant t. We denote :

- the thrust magnitude $u(t) = \|U(t)\|$, with $0 \le u \le u_{max}$
- the thrust direction $\vec{d}(t)$

The vehicle dynamics is represented by the first order ODE :

$$\dot{X}(t) = f[X(t), U(t), t] \Leftrightarrow \begin{cases} \dot{\vec{r}} = \vec{v} \\ \dot{\vec{v}} = \vec{g} + \dfrac{u.\vec{d}}{m} \\ \dot{m} = -\dfrac{u}{v_e} \end{cases} \quad (2)$$

where $v_e$ is the burned propellant exhaust velocity.

We denote $t_i^d$ and $t_j^a$ are the respective dates of departure from the debris i and arrival to the debris j. The vehicle state (position and velocity) must coincide :

- with the debris i state at the beginning of the transfer :  $Y(t_i^d) = Y_i(t_i^d)$
- with the debris j state at the end of the transfer :  $Y(t_j^a) = Y_j(t_j^a)$

The propellant consumption (or cost) $C_{ij}$ and duration $T_{ij}$ required for the transfer from the debris i to the debris j are respectively :

$$C_{ij}\left(t_i^d, t_j^a, X(t_i^d), X(t_j^a), [U_{ij}(t), t_i^d \leq t \leq t_j^a]\right) = \int_{t_i^d}^{t_j^a} \dfrac{u_{ij}(t)}{v_e} dt \quad (3)$$

and  $T_{ij} = t_j^a - t_i^d$ \quad (4)

The optimal control problem of the transfer from a debris i to a debris j is then formulated as :

$$\underset{t_i^d, t_j^a, U_{ij}(t)}{\text{Min}} C_{ij}\left(t_i^d, t_j^a, X(t_i^d), X(t_j^a), [U_{ij}(t), t_i^d \leq t \leq t_j^a]\right) \text{ s.t. } \begin{cases} Y(t_i^d) = Y_i(t_i^d) \\ Y(t_j^a) = Y_j(t_j^a) \\ \dot{X}(t) = f[X(t), U_{ij}(t), t] \text{ for } t_i^d \leq t \leq t_j^a \\ T_{ij} = t_j^a - t_i^d \leq T_{ij\,max} \end{cases} \quad (5)$$

The upper bound $T_{ij\,max}$ on the transfer duration is necessary in order to have a well posed problem[8,10]. The unknowns are the initial date $t_i^d$, the final date $t_j^a$, and the command law $\left[U_{ij}(t), t_i^d \leq t \leq t_j^a\right]$ which is a function of the time.

### B) Path problem

**The Travelling Salesman Problem**

The archetype of path problems is the Travelling Salesman Problem (TSP). The salesman has to visit successively N towns. Each town must be visited once and once only. The distance (or cost) from any town i to any other town j is $C_{ij}$ and the associated duration is $T_{ij}$ (Figure 2).

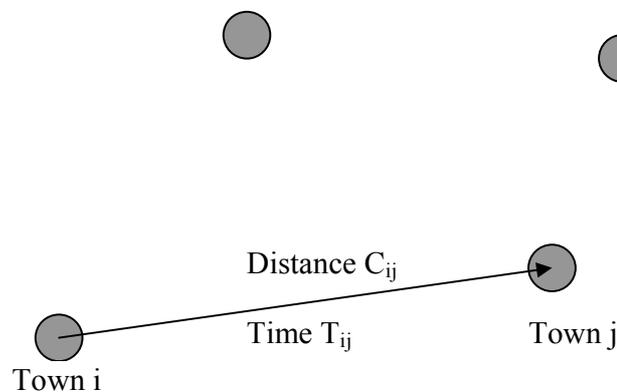

**Figure 2 : The Traveling Salesman Problem**

In an instance of the classical TSP, every town is linked to any other and the distances are constant. The goal is to minimize the total distance covered. There is no other constraint on the solution than to visit every town once and once only.

In order to formulate the SDC path problem, we start from the classical TSP with three additional features :

- Only n debris among the N must be visited
- There is a global duration constraint on the solution : $T < T_{max}$
- There is a minimum duration $T_{deorb}$ required to process a debris (capture, deorbitation, release). It corresponds to a minimum waiting phase once a debris has been reached before starting the next transfer.

**Path problem formulation**

In terms of graphs theory, the debris are the nodes and a transfer from a debris i to a debris j is represented by an edge from the node i to the node j. To write down the mathematical formulation of the path problem, we introduce several variables and related constraints[3].

Firstly we need to select the edges that will compose the path.

- $s_{ij}$ , i,j=1 to N is a binary selection variable for the edge (i-j) with i≠j:

    $s_{ij}$ = 1 if the edge is selected on the path

    $s_{ij}$ = 0 if not

    In order to select n debris amongst N, we must select n-1 edges : $\sum_{i,j} s_{ij} = n - 1$

    There are N(N-1) variables $s_{ij}$ and 1 related constraint.

Secondly we need to constrain the number of edges arriving and departing from each debris.

- $x_k$ , k=1 to N is the number of edges arriving to the node k :

    Since there will be at most one edge arriving to each debris, we can consider $x_k$ as a binary variable satisfying : $x_k = \sum_i s_{ik}$

There are N variables $x_k$ and N related constraints.

- $y_k$, k=1 to N is the number of edges departing from the node k :

  Since there will be at most one edge departing from each debris, we can consider $y_k$ as a binary variable satisfying : $y_k = \sum_j s_{kj}$

  There are N variables $y_k$ and N related constraints.

Thirdly we need to build a physical path wrt the chronology (increasing dates) and to forbid loops.

- $t_k^a$ and $t_k^d$, k=1 to N are the dates of arrival to and departure from the debris k

  The mission is assumed to start at $t_0=0$ and end at $t < T_{max}$ (1 year)

  $t_k^a$ and $t_k^d$ are therefore real variables representing comprised between 0 and $T_{max}$.

  The constraints related to the events chronology are :

  $$t_k^d - t_k^a \geq T_{deorb} \quad \text{(for the debris operations)}$$

  $$t_i^d - t_j^a + T_{max} \cdot s_{ij} \leq T_{max} - T_{min}$$

  $T_{min}$ represents a lower bound on the transfer duration in order to exclude unrealistic durations. It can for example be set to 1 day.

  This last constraint ensures that the dates are still increasing throughout the selected path. Indeed $t_i^d \leq t_j^a - T_{min}$ if the edge (i-j) is selected on the path. It therefore prevents from the solution from any loop.

  There are 2N variables $t_k^a$ and $t_k^d$ and N+N(N-1)=N² related constraints.

Fourthly we need to make the path connex in order to be physically feasible by the SDC vehicle.

- $z_k$ , k=1 to N is the product of $x_k$ and $y_k$ :    $z_k = x_k \cdot y_k$

    $z_k$ is therefore a binary variable such that

    $z_k = 1$ if there is one edge arriving and one edge departing from the debris k

    $z_k = 0$ else

    By adding the constraints : $z_k = x_k \cdot y_k$ and $\sum_k z_k = n - 2$ to the previous ones, we ensure that the path becomes connex (see below).

    There are N variables $z_k$ and N+1 related constraints.

The path connexity stems from the combination of the previous constraints :

- From :    $\sum_k x_k = \sum_{i,k} s_{ik} = n - 1$    (number of "arrival" debris)

    And    $\sum_k z_k = n - 2$    (number of "mid path" debris)

    We deduce that there is only one debris which is an arrival without departure (end of the mission).

- From :    $\sum_k y_k = \sum_{k,j} s_{kj} = n - 1$    (number of "departure" debris)

    And    $\sum_k z_k = n - 2$    (number of "mid path" debris)

    We deduce that there is only one debris which is a departure without arrival (beginning of the mission).

- From :    $t_i^d \leq t_j^a - T_{min}$   if $s_{ij} = 1$    (increasing dates along the path)

    And    $\sum_{i,j} s_{ij} = n - 1$    (number of edges)

And $\sum_{k} z_k = n - 2$ (number of "mid path" debris)

We deduce that the path is made of a n-1 consecutive edges without loops. There cannot be 2 separate paths as pictured on Figure 3.

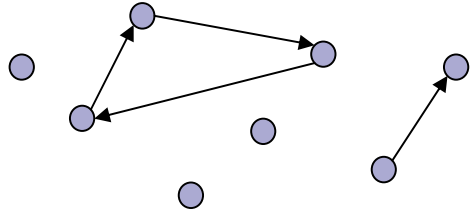

**Figure 3 : Non connex path with loop**

The SDC mission is composed of transfer phases between the successive selected debris and debris operations phases (capture, deorbiting, releasing). The respective costs and durations are :

- $C_{ij}$ and $T_{ij}$ for the transfer (i-j) with : $T_{ij} = t_j^a - t_i^d$

- $C_k$ and $T_k$ for the debris k operations with : $T_k = t_k^d - t_k^a$

In order to formulate the path total cost and duration, we define for each debris k a binary selection variable $s_k$. These N variables $s_k$ with one index k are attached to each debris, and must not be confused with the N.(N-1) variables $s_{ij}$ with two indexes i,j attached to each transfer. The variable $s_k$ is related to the binary variables $x_k$, $y_k$, $z_k$ previously defined by :

$s_k = x_k + y_k - z_k = x_k + y_k - x_k.y_k$

Thus : $s_k = 1$ if the debris is selected on the path (either $x_k = 1$ or $y_k = 1$)

$s_k = 0$ if not ($x_k = y_k = 0$)

The path total cost C and duration T are then :
$$\begin{cases} C = \sum_{i,j} s_{ij} C_{ij} + \sum_k s_k C_k \\ T = \sum_{i,j} s_{ij} T_{ij} + \sum_k s_k T_k \end{cases}$$

### C) Space Debris Collecting problem

The SDC problem is a non linear time dependent variant of the classical TSP. The main differences with the TSP come from the embedded transfer problems :

- The towns locations are replaced by the debris locations on their orbits (therefore varying with the time)

- There are an infinite number of possible trajectories to go from any debris i to any other debris j, depending on the departure date $t_i^d$ and arrival date $t_j^a$ and on the command law $[U_{ij}(t), t_i^d \leq t \leq t_j^a]$ applied during the transfer (Figure 4). Each possible trajectory requires thus a propellant consumption (cost) $C_{ij}(t_i^d, t_j^a, X(t_i^d), X(t_j^a), [U_{ij}(t), t_i^d \leq t \leq t_j^a])$ and a duration $T_{ij} = t_j^a - t_i^d$.

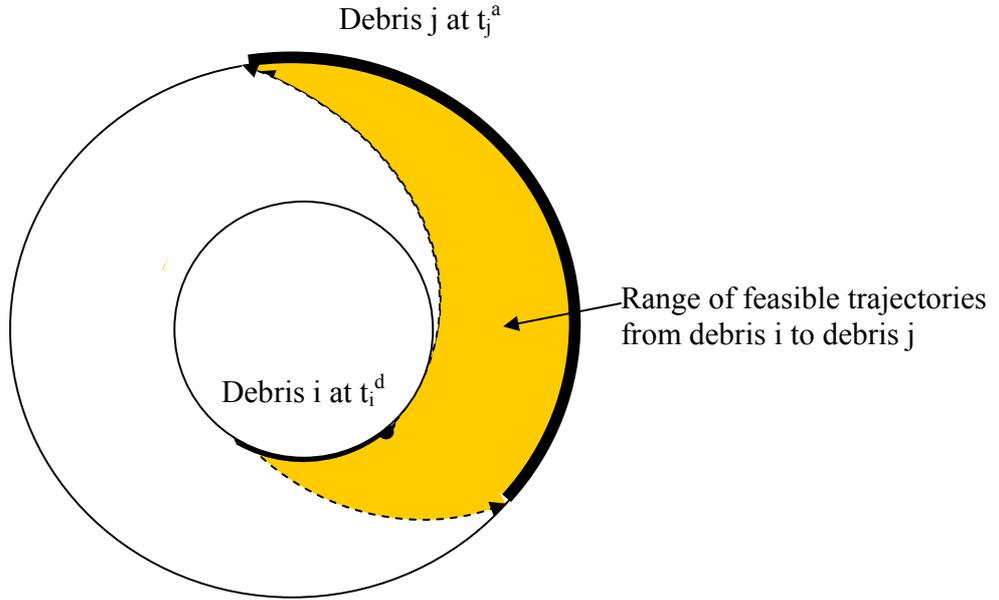

**Figure 4 : Trajectories between debris**

The SDC problem formulation is then :

$$\min_{s_{ij},x_k,y_k,z_k,s_k,t_k^a,t_k^d,X(t_k^d),X(t_k^a),U_{ij}(t)} C \quad \text{s.t.} \quad \begin{cases} S = n-1 \\ T \leq T_{max} \\ x_k = \sum_i s_{ik} \\ y_k = \sum_j s_{kj} \\ z_k = x_k \cdot y_k \\ s_k = x_k + y_k - y_k \\ \sum_k z_k = n-2 \\ t_k^d - t_k^a \geq T_{deorb} \\ t_i^d - t_j^a + T_{max} \cdot s_{ij} \leq T_{max} - T_{min} \\ Y(t_i^d) = Y_i(t_i^d) \\ Y(t_j^a) = Y_j(t_j^a) \\ \dot{X}(t) = f[X(t), U_{ij}(t), t], \quad \text{for} \quad t_i^d \leq t \leq t_j^a \end{cases} \quad (6)$$

where :

- N and n are respectively the total number of debris and the number of debris to deorbit

- S is the number of transfers : $\qquad S = \sum_{i,j} s_{ij}$

- C is the mission cost : $\qquad C = \sum_{i,j} s_{ij} C_{ij} + \sum_k s_k C_k$

- T is the mission duration : $\qquad T = \sum_{i,j} s_{ij} T_{ij} + \sum_k s_k T_k$

- Y is the location state vector : $\qquad X(t) = \begin{pmatrix} Y(t) \\ m(t) \end{pmatrix}$

- $C_{ij}$ is the cost of the transfer (i-j) : $\quad C_{ij}\left(t_i^d, t_j^a, X(t_i^d), X(t_j^a), [U_{ij}(t), t_i^d \leq t \leq t_j^a]\right) = \int_{t_i^d}^{t_j^a} \frac{u_{ij}(t)}{v_e} dt$

- $C_k$ is the cost of the debris k operations : $C_k\left(t_k^a, t_k^d, Y_k(t_k^a), Y_k(t_k^d)\right)$

- $T_{ij}$ is the duration of the transfer (i-j) : $\qquad T_{ij} = t_j^a - t_i^d$

- $T_k$ is the duration of the debris k operations : $\qquad T_k = t_k^d - t_k^a$

The mission cost and duration are assessed from the first debris selected on the path. It is assumed that a launcher has realized the rendezvous maneuvers in order to bring the SDC vehicle to this first debris. If this is not the case, the same formulation (Equation 6) applies by adding a fictitious debris numbered 0 corresponding to orbit of injection of the SDC vehicle. This fictitious debris is fixed as the start of the path by setting the related variables ($x_0=0$, $y_0=1$, $z_0=0$ and $s_0=1$). The mission assessment takes thus into account the transfer from this debris 0 to the first selected debris.

This formulation mixes :

- binary variables :   $s_{ij}, x_k, y_k, z_k, s_k$
- real variables :   $t_k^d, t_k^a, X(t_k^d), X(t_k^a)$
- functions :   $[U_{ij}(t), t_i^d \leq t \leq t_j^a]$

### D) Numerical methods

The SDC problem stated above is a mix of hard optimization problems :

- Several embedded functional optimization problems consisting in finding the optimal control law of a dynamic system
- A combinatorial optimization problem consisting in selecting the optimal path in a graph

The features of these two problems are recalled here after with an overview of the existing resolution methods.

**Functional optimization**

Functional optimization deals with infinite dimension problems, one at least of the unknowns being a function. For such problems, even simple, there is generally no systematic solution. For each instance of the problem, a numerical solution must be searched with iterative algorithms. The numerical methods split into two main categories[14,19] :

- Direct methods discretize the command law with time steps. The functional problem is transformed into a large size finite dimension problem for which NLP algorithms can be applied. The direct methods are relatively easy to initialize, but the convergence is generally slow and inaccurate.

- Indirect methods try to solve the infinite dimension problem using the PMP. A costate vector of the same dimension of the state vector is introduced, which obeys the Euler-Lagrange equations. The problem reduces to finding the initial values of the costate vector in order to match the optimality necessary conditions. A shooting method with a correct initialization can be used to solve these non linear equations. The indirect methods are very difficult to initialize, but the convergence is generally quick and accurate.

**Combinatorial optimization**

Combinatorial optimization deals with problems in which all or part of the unknowns are integer variables. The classification of combinatorial problems lies on the existence or not of polynomial algorithm, on one hand to solve the problem, on the other hand to check a solution[17].

The P-class includes problems with a known polynomial solving algorithm. The same algorithm can therefore be used to check the solution.

The NP-class includes problems with a known polynomial checking algorithm, but no known polynomial solving algorithm. The set of NP complete problems denoted NPC is formed by all NP problems that can be transformed in each other by a polynomial algorithm. If one day a polynomial solving algorithm was found for one problem of NPC, then any problem of NPC could also be solved polynomially and we would have NP=P.

The TSP and all its variants are NP complete (or NP-hard or NP-difficult). For these hard combinatorial problems, the resolution methods fall into three main categories[12,17] :

- Explicit enumeration of all the possible combinations ensures to find the exact solution i.e. the global optimum. The total number of arrangements of n debris amongst N is equal to $A_N^n = \frac{N!}{(N-n)!}$. Taking into account the constraints leads to formulate the problem with large numbers of binary variables. With m binary variables, the total number of combinations amounts to $2^m$. The explicit enumeration is practically restricted only to small size problem with a few ten variables. For m=50, the enumeration would take one year with a computer assessing one billion combinations per second.

- Implicit enumeration (Branch and Bound, Branch and Cut) consists in exploring the tree of all possible combinations with a cut-off of branches during the exploration[7,16]. The cut-off is based on an assessment of the best potential solution contained in a branch, and the comparison of this solution with a reference admissible solution. Likewise explicit enumeration, implicit enumeration yields the exact solution. The method efficiency is highly dependent on the branching strategy. Also the computation time grows exponentially with the problem size. These methods are therefore applicable to medium size problem (a few hundred variables).

- Approximate solutions can be found by stochastic programming methods like genetic algorithm, tabu search, simulated annealing…[9,12]. These methods limit a priori the number of combinations assessed and thus the computation time. The full combinations space is explored using an oriented random strategy. If the best solution found is judged unsatisfying, the search may be resumed with new tunings in order to improve the exploration strategy. There is generally no guarantee on the local optimality of the solution nor on its difference from the true global optimum. These methods are the only ones applicable for large size problems (more than thousands variables).

**Selected algorithms for the SDC problem**

The SDC problem features are :

- The medium size : 5 debris must be selected amongst lists of typically 10 debris. The number of variables and related constraints required to formulate the path problem ranges between 100 and 1000. Although the problem is NP-complete, this medium size favours the choice of an implicit enumeration method like Branch and Bound in order to get the exact solution.

- The difficulty due to the infinite dimension embedded control problems, which are themselves intrinsically hard in the general case. The cost of going from any debris to the other depends on the starting date and on the allocated duration. The mission overall duration being constrained to less than one year, there is a strong coupling between the optimal control laws of the successive transfers from one debris to the other.

To hope finding the global optimum, the cost and duration couplings between the successive transfers must be taken into account within the path problem resolution. This holds whatever the resolution method selected, approximate or exact. In order to get a tractable formulation, the embedded optimal control problem must first be simplified. The simplification aims at :

- reducing sufficiently the problem dimension so that it can be solved in reasonable computation times
- formulating the problem so that it can be solved by a robust and efficient algorithm as many times as needed

- keeping a realistic modelling wrt to physical trajectories.

In view of a Branch and Bound algorithm, a linear formulation of the problem is desirable, in order to ensure the required solving robustness and efficiency. The simplifications steps leading to a practical formulation that can be solved by a Branch and Bound algorithm consist in :

- Defining a generic transfer strategy in order to restrict the general optimal control problem and reduce the problem size

- Formulating the problem wrt the generic transfer strategy selected (transfer variables and constraints)

- Linearizing the resulting formulation around an initial solution in order to apply linear programming methods within the Branch and Bound algorithm.

Due to the linearization step, an iterative process must also be set up in order to update the linearization around the new solution, until the solution stabilizes. These successive simplifications steps detailed in the next sections.

## III TRANSFER STRATEGY

There is no known general transfer strategy that would be optimal whatever the initial and final orbits, the mission constraints and the vehicle capabilities. For the SDC mission it is nevertheless possible to define a specific transfer strategy that takes advantages of the mission particularities, particularly regarding the debris orbits and the mission duration.

### A) Debris orbits

For each debris, the orbital parameters are given at the mission starting date $t_0$ :

- a(t$_0$) = semi-major axis (m)
- e(t$_0$) = eccentricity (-)
- I(t$_0$) = inclination (deg)
- Ω(t$_0$) = right ascension of the ascending node (RAAN) (deg)
- ω(t$_0$) = argument of the perigee
- ν(t$_0$) = true anomaly

These parameters define the orbit shape (a,e), the orbit plane (I,Ω) and the location on the orbit (ω,ν).

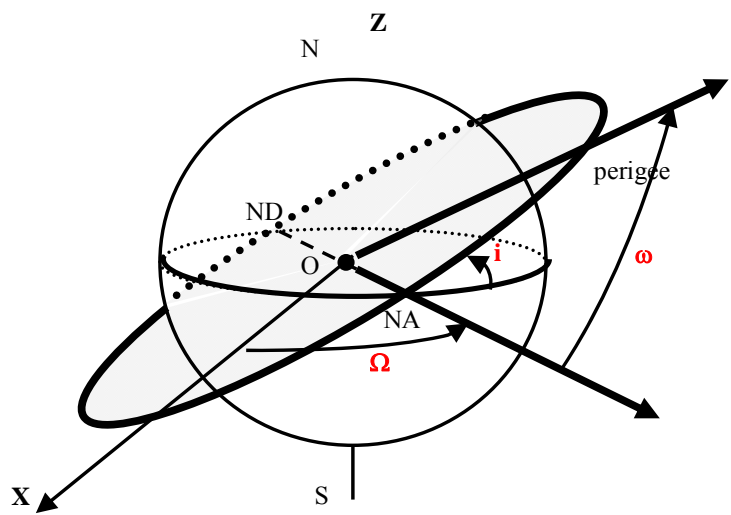

**Figure 5 : Orbital parameters**

The Space Debris Collecting mission aims at Low Earth Orbits (LEO) or Sun-Synchronous Orbits (SSO) old satellites. During their operational life the orbital parameters have been controlled to keep precise values matching the mission purposes, e.g. observation or telecommunication. Similar missions lead to very close operational orbits. For example typical SSO are circular at altitudes ranging from 600 to 1000 km, and inclination ranging from 96 to 100 deg. Similarly the launchers upper stages used to bring the satellites on their operational orbit have been left in the vicinity of these orbits, and constitute also targets for the SDC mission.

After their end of life, the satellites have been left uncontrolled and their orbits are subjected to perturbations (Earth gravitational perturbations, Sun and Moon attraction, atmospheric drag, pressure radiation, geomagnetic field, ...). For nearly circular LEO or SSO, the average effect of the perturbations on the orbit semi-major axis, the eccentricity and the inclination is very small : this is confirmed both by numerical simulations and by the observation of the debris orbits evolution[4]. A SDC mission targets therefore at series of abandoned spacecrafts moving on nearly circular orbits, at very close altitudes and inclinations.

In the altitude range [600 km – 1000 km] the main orbital perturbation is due to the Earth flattening, which adds the first zonal term $J_2$ in the Earth gravitational potential. The Earth equatorial bulge creates a torque on the debris orbit. The debris rotating on its orbit behaves as a gyroscope : the angular momentum rotates around the Earth polar axis causing a precession of the orbital plane and a secular drift of the node along the Equator (Figure 6).

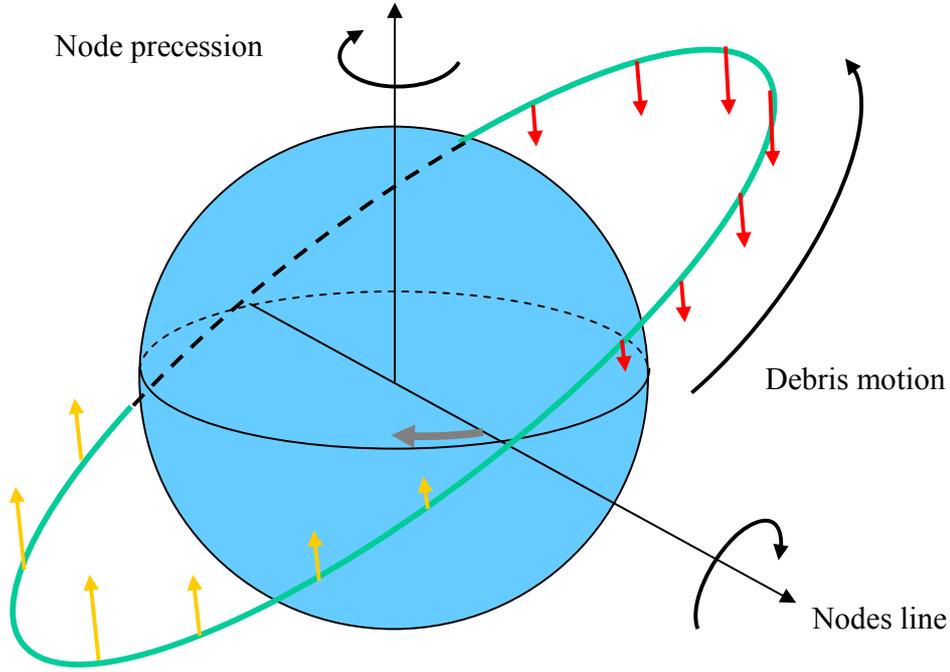

**Figure 6 : Orbit precession due to the Earth flattening**

The secular precession rate of the RAAN is expressed as[5]:

$$\dot{\Omega} = \frac{d\Omega}{dt} = -\frac{3}{2} J_2 \, n \left(\frac{R_T}{a}\right)^2 \frac{1}{(1-e^2)^2} \cos I = -C_{J2} \frac{\cos I}{a^{\frac{7}{2}}(1-e^2)^2} \quad (7)$$

where : - $R_T$ is the Earth equatorial radius (= 6378137 m)

- $J_2$ is the first zonal coefficient (= 1.086 $10^{-3}$), $C_{J2} = \frac{3}{2} J_2 \sqrt{\mu} \, R_T^{\,2} = 1.318 \, 10^{18}$

- $\mu$ is the Earth gravitational constant (= 3.986 $10^{14}$ m³/s²)

- n is the mean motion : $n = \sqrt{\frac{\mu}{a^3}}$

The other parameters a, e and I are not subjected to secular effects from the $J_2$ perturbation and they can be considered as constant throughout the SDC mission. In particular, the debris orbit remains nearly circular and the debris location on its orbit can be defined by the longitude from the ascending node L=ω+ν which replaces the two parameters ω and ν.

The RAAN precession rate which depends on a,e,I (Equation 7) is therefore constant :
$\dot{\Omega}(t) = \dot{\Omega}(t_0) = \dot{\Omega}_0$

and the RAAN is a linear function of the time : $\Omega(t) = \Omega(t_0) + \dot{\Omega}_0 \times (t - t_0)$

For a Sun-Synchronous Orbit the precession rate is adjusted to be equal to the Earth revolution rate around the Sun (360 deg in 1 year, i.e. 0.986 deg/day). The orbital plane makes a complete revolution around the polar axis in one year keeping a constant angle with the Sun direction, as illustrated on Figure 7.

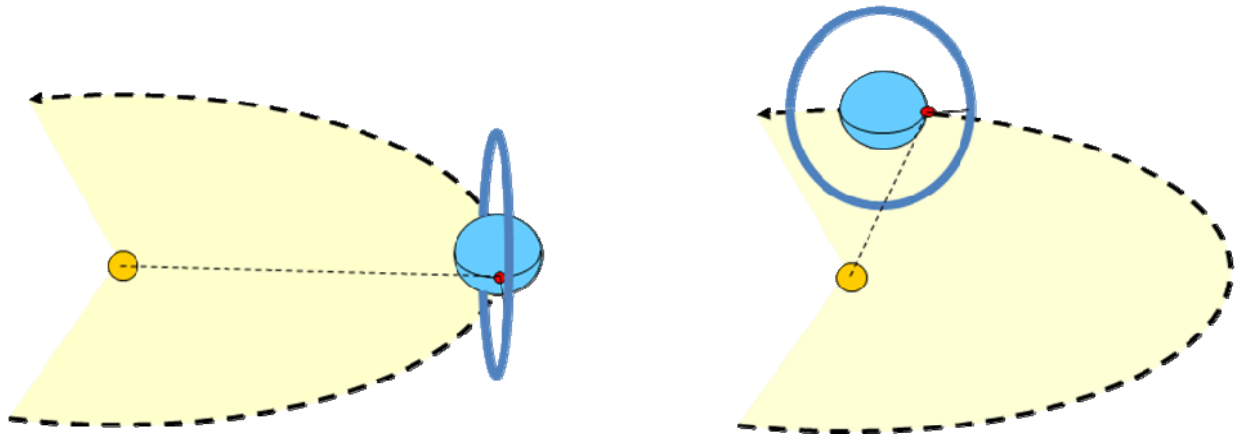

**Figure 7 : Sun-Synchronous Orbit**

Table 2 gives the range of values of the RAAN precession rates for near SSO debris.

| RAAN precession rate | | Semi major axis | |
|---|---|---|---|
| | | 7000 km | 7200 km |
| Inclination | 98 deg | 1.002 deg/day | 0.908 deg/day |
| | 99 deg | 1.126 deg/day | 1.020 deg/day |

**Table 2 : RAAN precession rates for near SSO debris**

In summary, the targeted debris orbital parameters can be split into :

- 3 constants : $a, e, I$ with $e \approx 0$

- 1 slow variable : $\Omega(t) = \Omega(t_0) + \dot{\Omega}_0 \times (t - t_0)$ with $\dot{\Omega}_0 = -C_{J2} \dfrac{\cos I}{a^{\frac{7}{2}}(1-e^2)^2}$

- 1 fast variable : $L(t) = L(t_0) + n \times (t - t_0)$ with $n = \sqrt{\dfrac{\mu}{a^3}}$

The transfer strategy is investigated through the following steps :

- Assess the performances (impulsive cost and duration) of a direct strategy
- Assess the performances (impulsive cost and duration) of a drift strategy
- Analyze the influence of the duration constraint.
- Analyze the influence of the thrust level.
- Define the generic strategy meeting the SDC mission specificities.

### B) Direct strategy

The debris orbits are quasi-circular at close altitudes and inclinations :

$$\begin{cases} 700 \text{ km} \leq H \leq 900 \text{ km} \\ 98 \text{ deg} \leq I \leq 99 \text{ deg} \end{cases} \text{ with } a = R_T + H \text{ for a circular orbit}$$

To gain some insight into the features of an optimal transfer strategy, we first assess the impulsive costs required for direct orbital parameters changes starting from a typical circular SSO (altitude H=800 km, inclination I=98.6 deg) :

- $\Delta H = \pm 100$ km $\Rightarrow$ $\Delta V = 50$ m/s (altitude correction)
- $\Delta I = \pm 1$ deg $\Rightarrow$ $\Delta V = 130$ m/s (inclination correction)
- $\Delta \Omega = \pm 1$ deg $\Rightarrow$ $\Delta V = 130$ m/s (RAAN correction)

    $\Delta \Omega = \pm 10$ deg $\Rightarrow$ $\Delta V = 1300$ m/s

The anomaly correction necessary for the rendezvous with the targeted debris is also assessed considering a one revolution Lambert manoeuvre :

- $\Delta L = \pm 10$ deg $\Rightarrow$ $\Delta V = 150$ m/s (anomaly correction)

    $\Delta L = -90$ deg $\Rightarrow$ $\Delta V = 1010$ m/s

    $\Delta L = +90$ deg $\Rightarrow$ $\Delta V = 1640$ m/s

All the debris altitudes and inclinations being very close, the changes of H and I to go from any debris to any other can be performed directly at moderate costs, typically less than 200 m/s, whatever the date. On the other hand, the RAAN differences changes with the time since each debris has a its own precession rate. The required RAAN correction may therefore take any values between -180 deg and +180 deg depending on the manoeuvre date. The impulsive cost increases roughly proportionally to the angle difference as illustrated in the above numerical example ($\Delta V \approx 130 \cdot |\Delta \Omega|$). The same remark applies to the anomaly correction if the manoeuvre date cannot be chosen freely.

A direct change strategy would lead therefore to prohibitive costs wrt to the RAAN and anomaly corrections.

### C) Drift strategy

An alternative strategy consists in waiting until the $J_2$ precession has nullified the RAAN difference. For two debris of respective orbital parameters $(a_1, e_1, I_1, \Omega_1)$ and $(a_2, e_2, I_2, \Omega_2)$ at an initial date $t_0$, the respective RAAN evolutions due to the $J_2$ precession are :

- $\Omega_1(t) = \Omega_1(t_0) + \dot{\Omega}_1 \times (t - t_0)$ with $\dot{\Omega}_1 = -C_{J2} \dfrac{\cos I_1}{a_1^{\frac{7}{2}}(1-e_1^2)^2}$

- $\Omega_2(t) = \Omega_2(t_0) + \dot{\Omega}_2 \times (t - t_0)$ with $\dot{\Omega}_2 = -C_{J2} \dfrac{\cos I_2}{a_2^{\frac{7}{2}}(1-e_2^2)^2}$

Considering debris 1 as the chaser and debris 2 as the target, the RAAN correction is performed when the chaser has caught-up the target : $\Omega_1(t) = \Omega_2(t)$. The required duration is:

$\Delta T = -\dfrac{\Omega_2 - \Omega_1}{\dot{\Omega}_2 - \dot{\Omega}_1} = -\dfrac{\Delta\Omega}{\Delta\dot{\Omega}}$. The value of $\Delta\Omega$ is to take modulo 360 deg in order to match the sign of $\Delta\dot{\Omega}$ and have a positive duration.

This purely waiting strategy is not time efficient for 2 reasons :

- The debris orbits are very close in terms of altitudes and inclinations, and the precession rates difference $\Delta\dot{\Omega}$ is therefore rather small (typically $|\Delta\dot{\Omega}| < 0.2$ deg/day from Table 2).

- The sign of $\Delta\dot{\Omega}$ may not match the optimal correction sense. This is illustrated on Figure 8 in the case of positive precession rate (retrograde orbits like SSO).

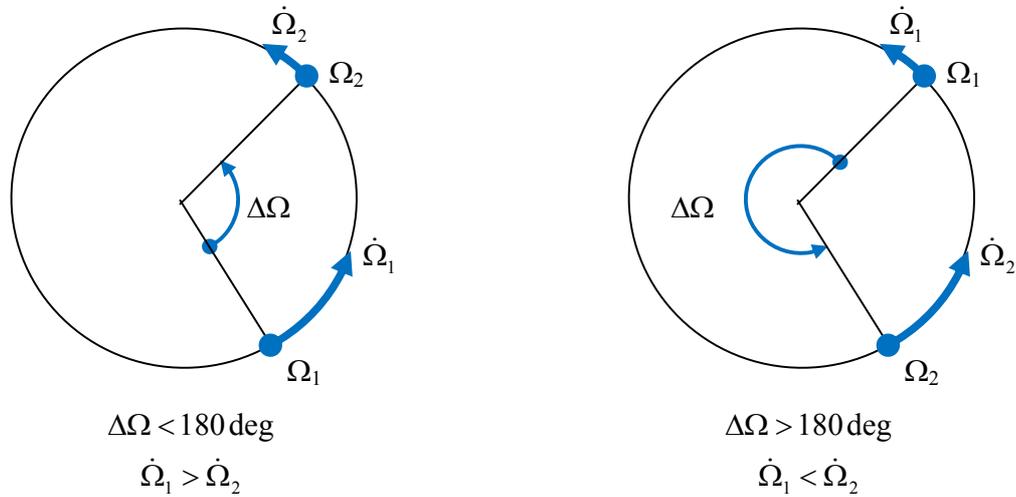

**Figure 8 : Forwards or backwards correction**

If the chaser is initially backwards ($\Delta\Omega < 180$ deg), it should try to overtake the target ($\dot{\Omega}_1 > \dot{\Omega}_2 \Rightarrow \Delta\dot{\Omega} < 0$). If the chaser is initially forwards ($\Delta\Omega > 180$ deg), it should try to w the target ($\dot{\Omega}_1 < \dot{\Omega}_2 \Rightarrow \Delta\dot{\Omega} < 0$).

In the worst cases (small $\Delta\dot{\Omega}$, wrong sense) waiting for a natural correction would take hundreds of days.

Similarly a waiting strategy for the anomaly correction is not time efficient, since it depends on the respective mean motions of the debris which can be very close.

In order to control the waiting duration, a more effective strategy consists in transferring the vehicle on an intermediate drift orbit where the RAAN correction will be speeded up. The choice of the drift orbit parameters allow to control both the sense of the correction (depending on whether the drift altitude is lower or higher than the chaser) and the correction speed (depending on the difference of the precession rates). A compromise has to be made between the cost of the additional transfer manoeuvres and the duration allocated to achieve the RAAN correction.

By the same way, a significant difference between the mean motions on the drift orbit and on the targeted debris orbit ensures that the anomaly phasing can be quickly (i.e. within a few revolutions) achieved whatever the initial anomaly difference. The additional cost due to the rendezvous constraint on the final orbit is then negligible, provided that the return transfer from the drift orbit to the debris orbit takes place at the adequate phasing opportunity once the RAAN correction is achieved.

### D) Duration constraint

The duration constraint is necessary to have a well posed problem else the optimal solution is in infinite time[8,10]. Qualitatively we can distinguish :

- A weak duration constraint if it makes the drift strategy is possible. The RAAN correction using the $J_2$ precession may then require several months, depending on the initial state. In that case the rendezvous cost can be considered as negligible wrt the orbit changes manoeuvres.

- A strong duration constraint if there is not enough time to use efficiently the $J_2$ precession for RAAN correction. In that case, the direct change strategy must be considered, taking directly into account the rendezvous constraint at the expense of higher mission costs. For example reference[11] presents a Lambert based strategy in order to deorbit 3 LEO debris within a few days.

### E) Thrust level

The term 'thrust level' will be used, although it would be more correct to speak of the acceleration level which reflects the orbit change capability of the vehicle. The way the transfer is optimized depends highly on the thrust level and on the transfer strategy :

- In the case of a high thrust engine, the manoeuvres durations can be considered as negligible wrt the coast phases durations, making the impulsive modelling adequate. It is then possible to analyze the transfer in a "patched-conics" manner, and to define a generic strategy based on Hohmann transfers (for the orbits changes) and Lambert transfers (for the rendezvous).

- In the case of a low thrust engine, the manoeuvres durations are no longer negligible. A global optimization is required taking into account the coupling between the successive phases of the transfer. In the case of the drift strategy, approximate solutions can be derived from the impulsive solution by taking into account the cost and duration penalty of low thrust manoeuvres. This approach is realistic provided that the manoeuvres durations remain compliant of the drift strategy. It this assumption is not satisfied, it is necessary to directly tackle the global transfer optimization problem through optimal control methods.

### F) Selected transfer strategy

The SDC mission start from the first selected debris and requires four successive transfers that must be realized within one year. The mean duration allocated for each transfer is 3 months. This leaves time to use the $J_2$ precession (weak duration constraint). The drift transfer strategy is therefore adequate for the SDC mission in order to minimize the mission overall cost.

In order to simplify the analysis and obtain a preliminary assessment of the mission cost, we assume in the following that the SDC vehicle is equipped with a high thrust engine, so that the impulsive approximation can be considered as valid. Extensions to low thrust engines is the subject of future work.

For each transfer from a debris 1 to a debris 2, the generic drift strategy is composed of 3 phases and it requires 4 manoeuvres :

- A first Hohmann transfer starting at the date $t_1$ to go from the debris 1 orbit to the intermediate drift orbit requires 2 apsidal manoeuvres $\Delta V_{P1}$ (perigee) and $\Delta V_{A1}$ (apogee)
- A waiting phase of duration $\Delta T_{12}$ on the drift orbit until the RAAN correction is completed
- A second Hohmann transfer starting at the date $t_2$ to go from the intermediate drift orbit to the debris 2 orbit requires 2 apsidal manoeuvres $\Delta V_{A2}$ (apogee) and $\Delta V_{P2}$ (perigee)

The transfer strategy is depicted on Figure 9 in the case of a drift orbit higher than the debris orbits. In the problem resolution there will be no a priori assumption regarding the relative altitudes of the 3 orbits. The drift orbit can be either below, or between or over the initial and final debris orbits.

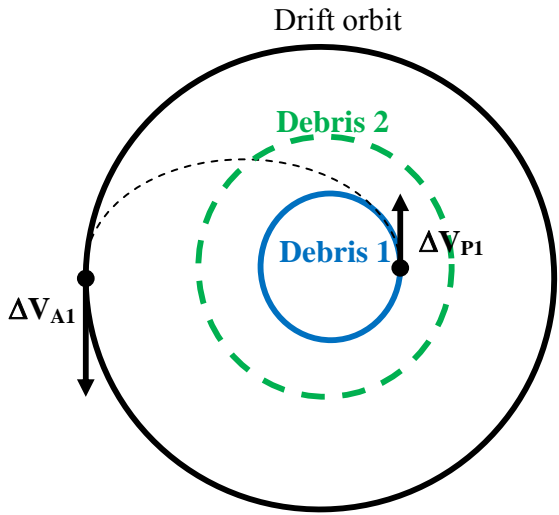 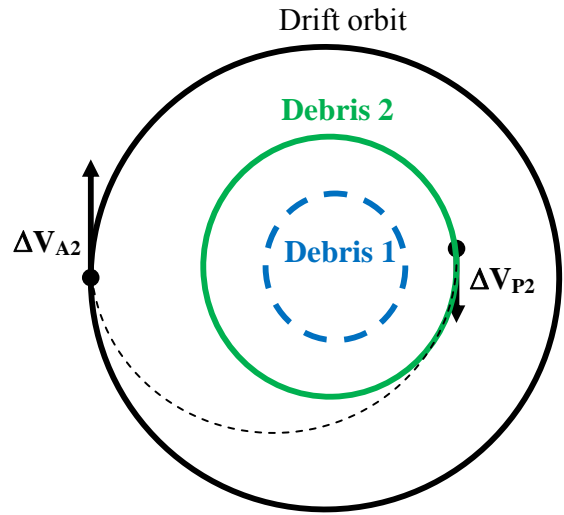

| Transfer from debris 1 orbit to drift orbit | Transfer from drift orbit to debris 2 orbit |

**Figure 9 : Transfer maneuvers and successive orbits**

The Hohmann transfers durations are considered as negligible (typically a half revolution, i.e. about 2h) wrt the drift duration (several weeks). The orbital parameters of the successive orbits are denoted :

- Debris 1 orbit departure at $t_1$ : $\quad a_1, e_1, I_1$ and $\Omega_1(t_1)$
- Drift orbit arrival at $t_1$ : $\quad a_d, e_d, I_d$ and $\Omega_d(t_1) = \Omega_1(t_1)$
- Drift orbit departure at $t_2 = t_1 + \Delta T_{12}$ : $\quad a_d, e_d, I_d$ and $\Omega_d(t_2) = \Omega_2(t_2)$
- Debris 1 orbit arrival at $t_2$ : $\quad a_2, e_2, I_2$ and $\Omega_2(t_2)$

In order to formulate the RAAN correction constraint we consider the RAAN evolution of the debris and of the SDC vehicle along their respective orbits with their respective precession rates since the mission starting date $t_0$ :

- Debris 1 :    $\Omega_1(t_1) = \Omega_1(t_0) + \dot{\Omega}_1 \times (t_1 - t_0)$

- Debris 2 :    $\Omega_2(t_2) = \Omega_2(t_0) + \dot{\Omega}_2 \times (t_2 - t_0)$

- Vehicle :    $\Omega(t_1) = \Omega_1(t_1)$    (start at $t_1$ from the debris 1 orbit)

    $\Omega(t_2) = \Omega_2(t_2)$    (arrival at $t_2$ on the debris 2 orbit)

    $\Omega(t_2) = \Omega(t_1) + \dot{\Omega}_d \times (t_2 - t_1)$    (coast from $t_1$ to $t_2$ on the drift orbit)

The drift orbit precession rate $\dot{\Omega}_d$ depends on the drift orbit parameters $a_d, e_d, I_d$ :

$$\dot{\Omega}_d = -C_{J2} \frac{\cos I_d}{a_d^{\frac{7}{2}}(1-e_d^2)^2} \tag{8}$$

The RAAN correction constraint can thus be written relatively to the mission starting date $t_0$ :

$$(\dot{\Omega}_2 - \dot{\Omega}_d) \times (t_2 - t_1) + (\dot{\Omega}_2 - \dot{\Omega}_1) \times (t_1 - t_0) + \Omega_2(t_0) - \Omega_1(t_0) = 0 \tag{9}$$

The duration required to complete the RAAN correction is given by :

$$t_2 - t_1 = \frac{\Omega_2(t_0) - \Omega_1(t_0) + (\dot{\Omega}_2 - \dot{\Omega}_1) \times (t_1 - t_0)}{\dot{\Omega}_d - \dot{\Omega}_2} \tag{10}$$

The anomaly rendezvous constraint is not considered since it can be satisfied with a negligible additional cost and duration by selecting the appropriate date of second transfer once the RAAN correction is achieved.

The overall cost of this generic transfer strategy is the sum of the 4 manoeuvres :

$$\Delta V = \Delta V_{P1} + \Delta V_{A1} + \Delta V_{A2} + \Delta V_{P2} \tag{11}$$

Each impulsive manoeuvre is assessed as the difference of the velocity vectors before and after the orbit change. The velocity modulus is given by the keplerian formula :

$$V = \sqrt{\mu\left(\frac{2}{r} - \frac{1}{a}\right)} \tag{12}$$

where   a is the orbit semi-major axis

r is the radius vector at the manoeuvre location.

In the case an inclination change is performed simultaneously with a shape change, the impulsive manoeuvre is assessed as the norm of the vectors difference (Figure 10) :

$$\Delta V = \left\|\vec{V}_a - \vec{V}_b\right\| = \sqrt{V_a^2 + V_b^2 - 2 V_a V_b \cos(i_b - i_a)} \tag{13}$$

where :   $\vec{V}_b$, $i_b$ are the velocity and inclination before the maneuver

$\vec{V}_a$, $i_a$ are the velocity and inclination after the maneuver

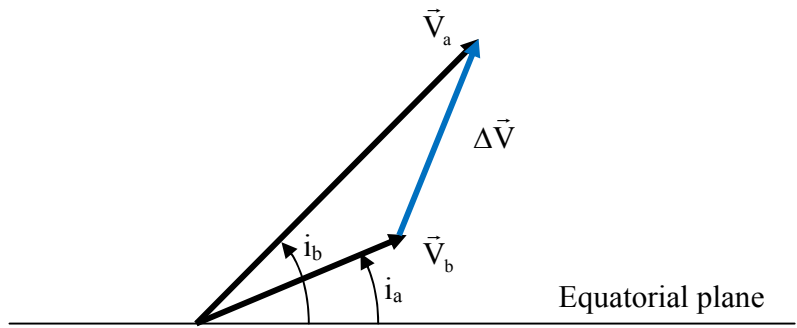

**Figure 10 : Simultaneous inclination and shape change**

**G) Problem formulation with the selected transfer strategy**

The parameters of the intermediate drift orbit related to each transfer (i-j) are denoted $a_{ij}$, $e_{ij}$, $I_{ij}$ with the corresponding precession rate :

$$\dot{\Omega}_{ij} = -C_{J2} \frac{\cos I_{ij}}{a_{ij}^{\frac{7}{2}}(1-e_{ij}^2)^2} \tag{14}$$

With the selected strategy composed of Hohmann transfers and drift waiting phases, the control problem (Equation 5) is simplified as follows :

- The command law $\left[U_{ij}(t), t_i^d \leq t \leq t_j^a\right]$ for each transfer (i-j) is replaced by the drift orbit parameters $a_{ij}$, $e_{ij}$, $I_{ij}$. There are no longer function unknowns and the problem becomes a non linear optimization problem of finite dimension.

- The dynamical constraint $\dot{X}(t) = f[X(t), U_{ij}(t), t]$, for $t_i^d \leq t \leq t_j^a$ disappears from the formulation since the trajectories are modelled by keplerian arcs (for the Hohmann transfers) with secular $J_2$ RAAN precession (for the drift orbits).

- The initial and final state constraints on $Y(t_i^d)$ and $Y(t_j^a)$ are directly taken into account in the transfer modelling from the initial debris orbit to the final one (the anomaly rendezvous constraint is neglected). They can thus be discarded from the problem formulation.

- The transfers costs $C_{ij}$ are measured by the required impulsive velocities $\Delta V_{ij}$, which depend only on the parameters of the initial, drift and final orbit. The mass can thus be discarded from the state vector.

- The transfers durations $T_{ij}$ also depend directly on the parameters of the initial, drift and final orbit :  $T_{ij} = t_j^a - t_i^d = \dfrac{\Omega_j(t_0) - \Omega_i(t_0) + (\dot\Omega_j - \dot\Omega_i) \times (t_i^d - t_0)}{\dot\Omega_{ij} - \dot\Omega_j}$

With these simplifications the SDC problem (Equation 6) becomes a mixed integer-real finite dimension problem whose formulation is :

$$\underset{s_{ij},x_k,y_k,z_k,s_k,t_k^a,t_k^d,a_{ij},e_{ij},I_{ij}}{\text{Min}} \; C \quad \text{s.t.} \quad \begin{cases} S = n - 1 \\ T \leq T_{max} \\ x_k = \sum_i s_{ik} \\ y_k = \sum_j s_{kj} \\ z_k = x_k \cdot y_k \\ s_k = x_k + y_k - z_k \\ \sum_k z_k = n - 2 \\ t_k^d - t_k^a \geq T_{deorb} \\ t_i^d - t_j^a + T_{max} \cdot s_{ij} \leq T_{max} - T_{min} \end{cases} \quad (15)$$

where :

- S is the number of transfers :  $S = \sum_{i,j} s_{ij}$
- C is the mission cost :  $C = \sum_{i,j} s_{ij} C_{ij} + \sum_k s_k C_k$
- T is the mission duration :  $T = \sum_{i,j} s_{ij} T_{ij} + \sum_k s_k T_k$

The cost $C_{ij}$ and duration $T_{ij}$ of the transfer (i-j) depend on the parameters of the initial orbit (debris i : $a_i, e_i, I_i, \Omega_i, \dot\Omega_i$), the final orbit (debris j : $a_j, e_j, I_j, \Omega_j, \dot\Omega_j$), the drift orbit ($a_{ij}, e_{ij}, I_{ij}, \Omega_{ij}, \dot\Omega_{ij}$). The duration depends also on the date of the transfer beginning $t_i^d$ :

$$\begin{cases} C_{ij} = \Delta V_{ij} \\ T_{ij} = t_j^a - t_i^d = \dfrac{\Omega_j(t_0) - \Omega_i(t_0) + \left(\dot{\Omega}_j - \dot{\Omega}_i\right) \times \left(t_i^d - t_0\right)}{\dot{\Omega}_{ij} - \dot{\Omega}_j} \end{cases}$$

The numbers of variables and constraints are :

- $N^2 + 3N$ binary variables :   $N(N-1)$ for the $s_{ij}$ and $4N$ for the $x_k, y_k, z_k, s_k$
- $3N^2 - N$ real variables :   $3N(N-1)$ for the $a_{ij}, e_{ij}, I_{ij}$ and $2N$ for the $t_k^d, t_k^a$
- $N^2 + 4N + 3$ constraints

### H) Further simplifications

It is possible to reduce further the problem dimension by investigating the influence of the drift orbit eccentricity and inclination on the transfer duration and cost.

**Drift orbit eccentricity**

The drift orbit eccentricity is a free parameter that influences both the cost of the Hohmann manoeuvres and the required duration in order to complete the RAAN correction.

We examine two extreme scenarios :

- The first scenario uses an elliptical drift orbit having the same perigee (or apogee) as the initial debris orbit. This scenario requires only three manoeuvres (one for the first transfer, two for the second transfer on the final debris orbit).
- The second scenario uses a circular drift orbit. This scenario requires four manoeuvres (two for the first transfer, two for the second transfer) as depicted on Figure 9.

In order to compare the two scenarios, we consider a representative numerical example with the initial and final debris on the same SSO at the altitude of 800 km and inclination of 98.6 deg. We assume also that the drift orbit altitude is comprised between 600 km and 1000 km.

The plots of Figure 11 show for the two scenarios the manoeuvres costs depending on the drift orbit altitude (or perigee or apogee in the elliptical scenario) and the precession duration required for a RAAN correction of one degree.

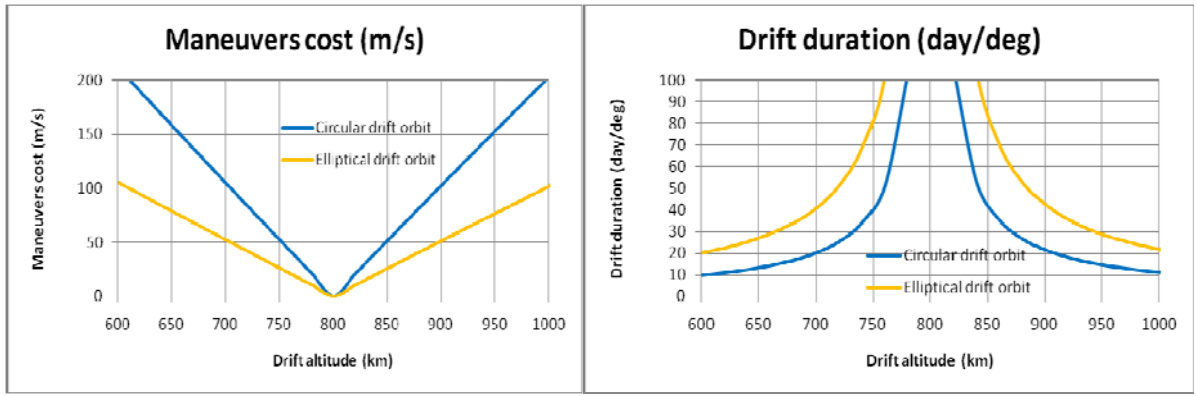

**Figure 11 : Elliptical vs. circular drift orbit**

These plots show that the fastest RAAN correction rate that can be hoped with the elliptical scenario is about 20 day/deg, with a related manoeuvres costs of 100 m/s. This means that a single RAAN correction of 10 deg could not be achieved in less than 200 days. This strategy risks therefore not being compliant with the SDC mission global duration of one year.

On the other hand the circular drift orbit scenario allows a doubling of the RAAN correction rate, but at the expenses of higher impulsive costs.

This numerical example shows that considering elliptical drift orbits can help reduce the mission cost, but makes the solution more sensitive to the duration constraint.

For a first approach, we will consider only circular drift orbits and fix the drift eccentricities $e_{ij}$ at zero. This assumption is the most conservative wrt the duration constraint, since it maximizes the RAAN correction rates of each transfer. It can later be relaxed when post-optimizing the trajectory (once the path is selected) in order to reduce the mission cost.

**Drift orbit inclination**

Regarding the drift orbit inclination a similar analysis can be made as for the eccentricity.

The respective inclinations on the initial and final debris orbits are $I_1$ and $I_2$. In order to minimize the overall transfer cost, the inclination change has to be performed simultaneously with one of the four manoeuvres of the transfer.

From the impulsive formula (Equation 13) : $\Delta V = \|\vec{V}_a - \vec{V}_b\| = \sqrt{V_a^2 + V_b^2 - 2V_a V_b \cos(i_b - i_a)}$ the minimum coupling cost is achieved for the minimum product of the velocities modulus $V_a V_b$.

For the SDC mission, the targeted debris are moving on nearly circular orbits at close altitudes. Assuming that the initial and the final debris altitudes are at the same altitudes $Z_1 = Z_2$ with respective inclinations $I_1$ and $I_2$, we can see that there are two cost equivalent opportunities for the inclination change (Figure 12). These opportunities correspond to the apogee manoeuvres :

- If the drift orbit is above the debris orbits, these opportunities are at the $2^{nd}$ and the $3^{rd}$ manoeuvres
- If the drift orbit is under the debris orbits, these opportunities are at the $1^{st}$ and the $4^{rd}$ manoeuvres

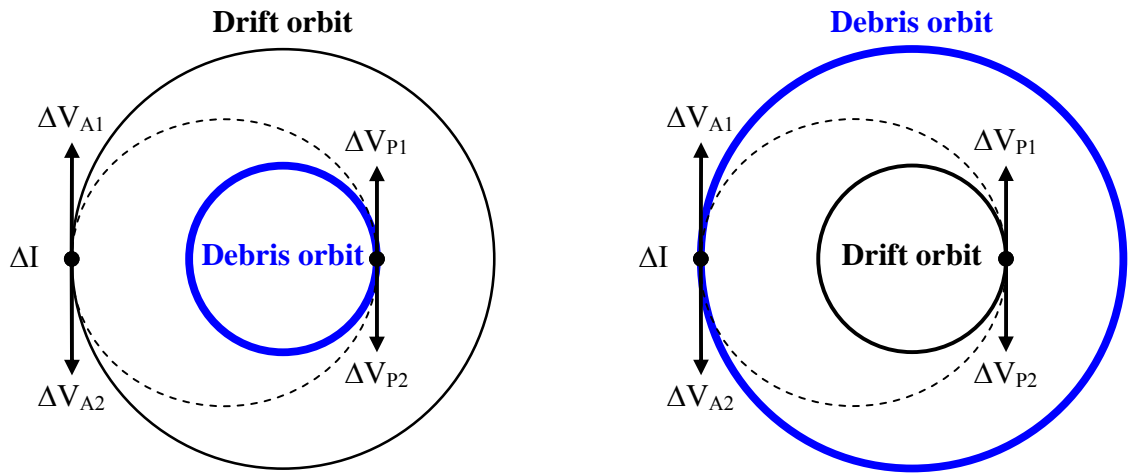

Drift orbit above the debris orbits        Drift orbit under the debris orbits

**Figure 12 : Optimal inclination change opportunity**

It is thus always possible to perform the inclination change either before or after the drift phase, without cost penalty. The criterion of choice is then to speed up the RAAN correction by maximizing the precession rate difference between the drift orbit and the final debris orbit.

For retrograde orbits (I > 90 deg), the precession rate is positive : $\dot{\Omega} = -C_{J2} \dfrac{\cos I}{a^{\frac{7}{2}}(1-e^2)^2} > 0$ and its variation with the semi-major axis and inclination are : $\begin{cases} \dfrac{\partial \dot{\Omega}}{\partial a} < 0 \\ \dfrac{\partial \dot{\Omega}}{\partial I} > 0 \end{cases}$

The optimal choice of the drift orbit inclination depending on the respective altitudes and inclinations is summarized in Table 3 :

| Drift inclination choice | $Z_{drift} < Z_2$ | $Z_{drift} > Z_2$ |
|---|---|---|
| $I_1 < I_2$ | $I_{drift} = Max(I_1, I_2) = I_2$ | $I_{drift} = Min(I_1, I_2) = I_1$ |
| $I_1 > I_2$ | $I_{drift} = Max(I_1, I_2) = I_1$ | $I_{drift} = Min(I_1, I_2) = I_2$ |

**Table 3 : Optimal drift inclination vs. RAAN correction duration**

For a first approach, we will fix the drift inclinations $I_{ij}$ at their optimal values wrt RAAN correction duration. In the practical context of the SDC mission this assumption is quasi-optimal in terms of cost as long as the debris move at very close altitudes. If not, the inclination change opportunities would not be equivalent and a compromise should to be made between minimizing the transfer cost and minimizing the RAAN correction duration.

**Debris operations**

In addition to the previous assumptions regarding the drift orbit parameters, we assume that :

- the durations $T_k$ for the debris operations are fixed ($T_k = T_{deorb}$) and identical for all the debris
- the minimum duration transfer $T_{min}$ is fixed to zero (it plays indeed a redundant role with $T_{deorb}$ in the problem formulation).
- the costs $C_k$ for the debris operations are constant and depend only on the debris orbits (cost of the rendezvous and deorbitation maneuvers)
- the vehicle leaves each debris as soon as possible once the deorbiting operations are completed : $t_k^d = t_k^a + T_{deorb}$.

Delaying the departure is indeed a loss of time : it is preferable to go as soon as possible on the next drift orbit in order to speed up the RAAN correction. It must also be noted that a drift orbit identical to the initial debris orbit is a possible solution, so that the above assumption fixing the departure date should in fact be close to the optimal solution.

The arrival date on the debris k will then be noted without superscript : $t_k^a = t_k$.

### I) Problem simplified formulation

With these simplifying assumptions, the SDC problem formulation (Equation 15) becomes :

$$\underset{s_{ij}, x_k, y_k, z_k, s_k, t_k, a_{ij}}{\text{Min}} \quad C \quad \text{s.t.} \quad \begin{cases} S = n - 1 \\ T \leq T_{max} \\ x_k = \sum_i s_{ik} \\ y_k = \sum_j s_{kj} \\ z_k = x_k \cdot y_k \\ s_k = x_k + y_k - z_k \\ \sum_k z_k = n - 2 \\ t_i - t_j + T_{max} \cdot s_{ij} \leq T_{max} - T_{deorb} \end{cases} \quad (16)$$

where :

- S is the number of transfers :  $\quad S = \sum_{i,j} s_{ij}$

- C is the mission cost :  $\quad C = \sum_{i,j} s_{ij} C_{ij} + \sum_k s_k C_k$

- T is the mission duration :  $\quad T = \sum_{i,j} s_{ij} T_{ij} + \sum_k s_k T_k$

$$\begin{cases} C_{ij}(a_{ij}) = \Delta V_{ij}(a_{ij}) \\ T_{ij}(a_{ij}, t_i) = t_j - t_i = \dfrac{\Omega_j(t_0) - \Omega_i(t_0) + (\dot{\Omega}_j - \dot{\Omega}_i) \times (t_i - t_0)}{\dot{\Omega}_{ij}(a_{ij}) - \dot{\Omega}_j} \end{cases}$$

The numbers of variables and constraints are :

- N² + 3N binary variables :    N(N-1) for the $s_{ij}$ and 4N for the $x_k, y_k, z_k, s_k$
- N² real variables :    N(N-1) for the $a_{ij}$ and N for the $t_k$
- N² + 3N + 3 constraints

## IV    PRACTICAL RESOLUTION

The above simplifications have removed the function unknowns and led to a finite reduced dimension problem. But the problem is still non linear, and therefore not adapted to a Branch and Bound approach. Indeed we will need to solve repeatedly instances of this problem, throughout the tree exploration. In order to guarantee both the global optimum and the time efficiency, every problem instance must be solved reliably and quickly.

It is very difficult, if not impossible, to ensure the required robustness (guaranteed convergence whatever the problem data) for a non linear problems. Such problems present generally local minima causing NLP algorithms to be sensitive to the initialization. Also it is nearly impossible to find algorithmic settings robust whatever the problem. For these reasons a non linear formulation cannot offer a sufficient guarantee of convergence and a user's control would be systematically necessary to check the solution optimality.

In order to get the required convergence guarantee and time efficiency, the most effective approach consists in linearizing the problem. The advantages are the elimination of local minima and the possibility to use reliable linear programming methods. The drawback is that the solution found is only valid in the vicinity of the starting point. Therefore the linearization and resolution must be iterated within an iterative process until the solution stabilizes.

### A) Problem linearization

The non linear terms in the above formulation (Equation 16) are :

- The transfers costs : $C_{ij}(a_{ij})$
- The transfers durations : $T_{ij}(a_{ij}, t_i)$
- The variables products : $(s_{ij}C_{ij}), (s_{ij}T_{ij}), (x_k y_k)$

**Transfer cost and duration linearization**

The cost and duration are linearized for each transfer around reference values of the drift orbit semi-major axis $a_{ij}$ and of the transfer starting date $t_i$. In order to ease the explanation of the linearization process, we denote :

- $a_d$ and $t_d$ the reference values of the semi major axis and the starting date
- a and t the actual values
- $\alpha$ and $\tau$ the differences between the actual and the reference values : $\begin{cases} a = a_d + \alpha \\ t = t_d + \tau \end{cases}$
- $[a_{min}; a_{max}]$ and $[t_{min}; t_{max}]$ the intervals in which the linearization is considered as valid. This notion of validity will be discussed when detailing the iterative process. The corresponding intervals on the variables $\alpha$ and $\tau$ are $[\alpha_{min}; \alpha_{max}]$ and $[\tau_{min}; \tau_{max}]$.

The non linear functions C(a) and T(a,t) depending on a and t are approximated in the intervals $[a_{min}; a_{max}]$ and $[t_{min}; t_{max}]$ by linear functions $C_L(\alpha)$ and $T_L(\alpha,\tau)$ depending on the differences of a and t with the reference values $a_d$ and $t_d$. These linear functions are built by a first linearization wrt the a variable, then by adding the time derivative for the duration function $T_L$.

$$\begin{cases} C_L(\alpha) = C(a_{min}) + \frac{\partial C}{\partial a}(\alpha - \alpha_{min}) \\ T_L(\alpha, \tau) = T(a_{min}, t_d) + \frac{\partial T}{\partial a}(\alpha - \alpha_{min}) + \frac{\partial T}{\partial t}\tau \end{cases} \quad (17)$$

The cost C and the duration T partial derivatives are approximated in the linearization intervals by secant formulae :

$$\begin{cases} \frac{\partial C}{\partial a} \approx \frac{C(a_{max}) - C(a_{min})}{a_{max} - a_{min}} \\ \frac{\partial T}{\partial a} \approx \frac{T(a_{max}, t_d) - T(a_{min}, t_d)}{a_{max} - a_{min}} & \text{assessed at } t = t_d \\ \frac{\partial T}{\partial t} \approx \frac{T(a_d, t_{max}) - T(a_d, t_{min})}{t_{max} - t_{min}} & \text{assessed at } a = a_d \end{cases} \quad (18)$$

This linearization choice for the duration function T is driven by the function shape. Denoting the initial and final debris with indexes 1 and 2, the duration function is :

$$T_{12} = \frac{\Omega_2(t_0) - \Omega_1(t_0) + (\dot{\Omega}_2 - \dot{\Omega}_1) \times (t_d - t_0)}{\dot{\Omega}_d(a_d) - \dot{\Omega}_2}$$

This function is linear wrt $t_d$ and highly nonlinear wrt $a_d$ with an asymptote at $a=a_2$ (same precession rate, and thus infinite duration) and an upwards concavity apart the asymptote (Figure 13).

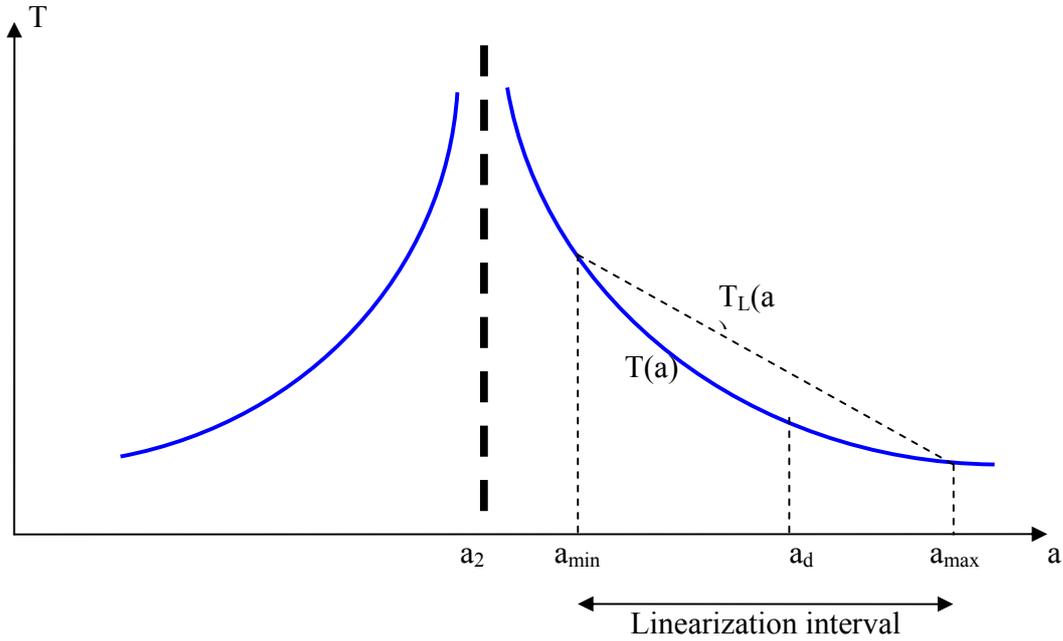

**Figure 13 : Duration function linear approximation**

The first linearization terms $T(a_{min}, t_d) + \frac{\partial T}{\partial a}(\alpha - \alpha_{min})$ yield an upper approximation of the true duration at the reference date $t_d$, while the last term $\frac{\partial T}{\partial t}\tau$ accounts for the linear dependence on the starting date.

The choice of an upper approximation ensures that the linearized solution overestimates the true duration of each transfer. The linearized solution is therefore sub-optimal wrt to the global duration constraint. The way to converge on the optimal solution within the iteration process will consist in tightening the intervals $[\alpha_{min}; \alpha_{max}]$ and $[\tau_{min}; \tau_{max}]$ from one iteration to the other after the linearized solution stabilizes.

**Alternative linearization approaches**

Two alternative linearization approaches wrt the semi-major axis variable have been envisioned.

The first one consists in considering directly the tangent at the linearization point instead of the secant on the linearization interval : $T_L(\alpha, \tau) = T(a_d, t_d) + \frac{\partial T}{\partial a}(a_d, t_d).\alpha + \frac{\partial T}{\partial t}(a_d, t_d).\tau$

This linearized duration function underestimates the true duration (the tangent is below the curve). The mission real duration corresponding to the linearized solutions is therefore systematically higher than 1 year. The intermediate iterations yield never feasible solutions wrt that constraint until the convergence is achieved. This first approach has proved practically difficult to control within the iteration process.

The second approach aims at meeting more accurately the true duration constraint at every iteration. It consists in approximating the duration function by 2 half segments as pictured on Figure 14 :

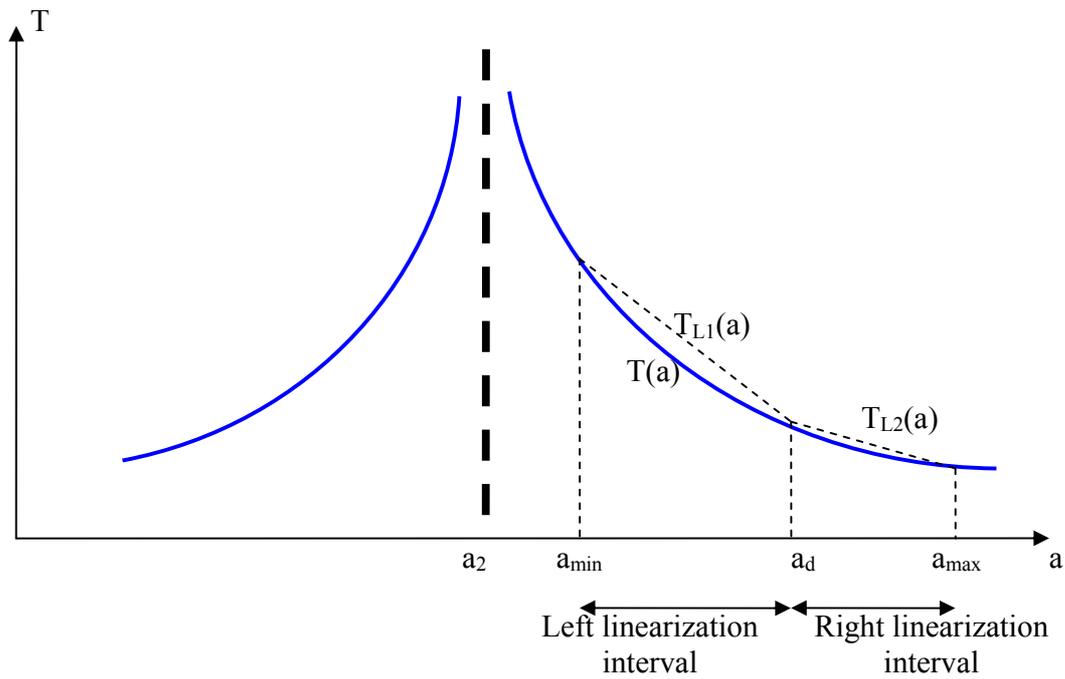

**Figure 14 : Duration function approximation with two half segments**

This approach makes necessary to take into account the sign of $\alpha_{ij}$ when linearizing the duration of the transfer (i-j) and it leads to an increase of the linearized problem dimension. The time of solving this larger dimension problem can nevertheless be balanced by a reduction of the number of iterations necessary to meet accurately the global duration constraint. The choice must be made case by case, depending on the iterations behaviour and on the computation time observed.

**Variables products linearization**

A product of a binary variable $x \in \{0;1\}$ by a real variable $y \in [y_{min;}, y_{max;}]$ is linearized by introducing one additional real variable z representing the product (x.y) and four additional constraints[3] :

$$\begin{cases} z \geq x.y_{min;} \\ z \leq x.y_{max;} \\ z \leq y - (1-x).y_{min;} \\ z \geq y - (1-x).y_{max;} \end{cases} \quad (19)$$

The variable z is then used in the problem formulation in order to replace everywhere the products (x.y). This transformation is applied to all the products $(s_{ij}.\alpha_{ij})$, $(s_{ij}.\tau_i)$ and $(x_k.y_k)$, adding thus $2N^2-N$ real variables and $8N^2-4N$ constraints.

**Linearized problem**

The linearized problem formulation is obtained by successively :

- Choosing for each transfer (i-j) reference values of $a_{ij}$ and $t_i$
- Replacing the true cost $C_{ij}$ and duration $T_{ij}$ by their respective linear approximations $C_{Lij}$ and $T_{Lij}$ (Equation 18),
- Replacing all the products by the associated variables and constraints (Equation 19).

We denote $C_{Lij}$ and $T_{Lij}$ the linearized cost and duration functions with their respective linearization coefficients $(C_{0ij}, C_{1ij})$ and $(T_{0ij}, T_{1ij}, T_{2ij})$ :

$$\begin{cases} C_{Lij}(\alpha_{ij}) = C_{0ij} + C_{1ij}\alpha_{ij} \\ T_{Lij}(\alpha, \tau) = T_{0ij} + T_{1ij}\alpha_{ij} + T_{2ij}\tau_{ij} \end{cases} \quad (20)$$

The products linearization variables are denoted :

$$\begin{cases} p_{ij} = s_{ij}\alpha_{ij} \\ q_{ij} = s_{ij}\tau_{ij} \end{cases} \tag{21}$$

The linearized formulation of the SDC problem (Equation 16) becomes :

$$\min_{s_{ij}, x_k, y_k, z_k, s_k, \alpha_{ij}, \tau_i, p_{ij}, q_{ij}} C_L \quad \text{s.t.} \quad \begin{cases} \sum_{i,j} s_{ij} = n-1 \\ \sum_k z_k = n-2 \\ x_k = \sum_i s_{ik} \\ y_k = \sum_j s_{kj} \\ s_k = x_k + y_k - z_k \\ \tau_i - \tau_j + T_{max}.s_{ij} \leq T_{max} - T_{deorb} \\ T_L \leq T_{max} \\ + \text{products linearization constraints for } z_k, p_{ij}, q_{ij} \end{cases} \tag{22}$$

where :

- $C_L$ is the linearized cost :  $\quad C_L = \sum_{i,j} s_{ij} C_{0ij} + \sum_{i,j} p_{ij} C_{1ij} + \sum_k s_k C_k$

- $T_L$ is the linearized duration :  $\quad T_L = \sum_{i,j} s_{ij} T_{0ij} + \sum_{i,j} p_{ij} T_{1ij} + \sum_{i,j} q_{ij} T_{2ij} + \sum_k s_k T_k$

The numbers of variables and constraints of the linearized problem are :

- $N^2 + 3N$ binary variables :  $\quad$ N(N-1) for the $s_{ij}$ and 4N for the $x_k, y_k, z_k, s_k$

- $3N^2 - 2N$ real variables :  $\quad$ 3N(N-1) for the $\alpha_{ij}, p_{ij}, q_{ij}$ and N for the $\tau_i$

- $9N^2 - 2N + 3$ constraints + variables bounds constraints.

### B) Initialization

The initialization goals are:

- To pre-optimize the parameters of the drift orbit for each transfer in order to start with an already good solution and limit the number of iterations

- To detect the unfeasible or too expensive transfers in order to eliminate them of the possible paths and thus reduce the problem dimension.

**Pre-optimization problem**

Considering a single transfer from a debris 1 to a debris 2, we denote as before the successive orbital parameters :

- Debris 1 orbit departure at $t_1$ :  $a_1, e_1, I_1$ and $\Omega_1(t_1)$
- Drift orbit arrival at $t_1$ :  $a_d, e_d, I_d$ and $\Omega_d(t_1) = \Omega_1(t_1)$
- Drift orbit departure at $t_2 = t_1 + \Delta T_{12}$ :  $a_d, e_d, I_d$ and $\Omega_d(t_2) = \Omega_2(t_2)$
- Debris 1 orbit arrival at $t_2$ :  $a_2, e_2, I_2$ and $\Omega_2(t_2)$

The RAAN correction constraint relatively to the mission starting date $t_0$ is :

$$(\dot{\Omega}_2 - \dot{\Omega}_d) \times (t_2 - t_1) + (\dot{\Omega}_2 - \dot{\Omega}_1) \times (t_1 - t_0) + \Omega_2(t_0) - \Omega_1(t_0) = 0$$

with the drift orbit precession rate $\dot{\Omega}_d$ : $\dot{\Omega}_d = -C_{J2} \dfrac{\cos I_d}{a_d^{\frac{7}{2}}(1-e_d^2)^2}$

The duration required to complete the RAAN correction is given by :

$$T_{12} = t_2 - t_1 = \dfrac{\Omega_2(t_0) - \Omega_1(t_0) + (\dot{\Omega}_2 - \dot{\Omega}_1) \times (t_1 - t_0)}{\dot{\Omega}_d - \dot{\Omega}_2}$$

The transfer pre-optimization consists in minimizing the transfer cost $\Delta V_{12}$ in less than a given duration $T_0$ : $\min\limits_{a_d, I_d, t_1, t_2} \Delta V_{12}$  s.t.  $T_{12} \leq T_0$

**Duration constraint**

The duration inequality constraint is active in most cases. Indeed shorter durations require higher precession rates differences and more expensive drift orbits. The constraint is inactive only if the debris precession rates are sufficiently different so that the RAAN correction is naturally completed within the prescribed duration. In that case the drift can occur on the initial orbit.

The SDC mission is composed of 4 transfers than must be completed is less than one year. The average duration per transfer is thus 3 months. Setting $T_0$ to 3 months at the pre-optimization step is in fact not robust. Indeed the linearized duration function $T_L$ used in the linear is an overestimate of the true duration in the linearization interval and this would make the linearized problem unfeasible wrt to the duration constraint.

The transfer maximal duration $T_0$ must therefore be set to strictly less than 3 months (e.g. 2 months), in order to keep a margin wrt the linearization error and have a feasible initialization of the linearized problem. The linearized solution will tend progressively to saturate the global duration constraint throughout the iterations.

**Solutions diagnosis and linearization**

For each transfer, the pre-optimization diagnoses if the transfer is unfeasible in the prescribed duration $T_0$ or more expensive than a prescribed cost threshold $\Delta V_{max}$. Such a transfer (i-j) is eliminated by fixing the corresponding selection variable $s_{ij}$ to zero. For a feasible transfer, the pre-optimization yields values of the semi-major axis $a_d$, the inclination $I_d$ and the initial date $t_1$. These values are taken as starting linearization values, with adapted linearization intervals. The linearization intervals $[\alpha_{min}; \alpha_{max}]$ and $[\tau_{min}; \tau_{max}]$ must be compliant with :

- The drift altitude bounds : $a_{min} < a_d < a_{max}$ .

  For SSO debris, the operational altitude range is [400 km ; 1200 km].

- The mission initial and final dates : $\begin{cases} t_0 < t_1 < t_0 + T_{max} \\ t_0 < t_2 < t_0 + T_{max} \end{cases}$ .

In order to have a valid linearized modelling of the transfer duration, the drift altitude interval must lies entirely on the same side of the targeted debris altitude. Indeed the transfer duration function has an asymptote for $a_d=a_2$ and $I_d=I_2$ (drift orbit identical to target orbit) and the linearization is subject to growing errors when one of the bounds $(a_d+\alpha_{min})$ or $(a_d+\alpha_{max})$ approaches the limit $a_d=a_2$ (Figure 13). This holds even if $I_d$ is not strictly equal to $I_2$, since all debris and drift orbits are in a narrow inclination range.

For each transfer, the side chosen for $a_d$ (lower or higher than $a_2$) at the initialization step is thus definitive for the subsequent iterations, since the linearized solution is bounded on one side of the asymptote. Consequently the selected inclination (either $I_d=I_1$ or $I_d=I_2$) is kept unchanged throughout the iterations.

These initial assumptions about the drift altitudes and inclinations are re-examined at the end of the iterations, in order to check their validity wrt to the optimal path found. Since all orbits are close on practical cases (debris orbits and drift orbits) these preliminary choices are generally optimal at the first attempt.

### C) Iteration process

We denote with a superscript $^{(k)}$ the drift orbit parameters at the $k^{th}$ iteration. The problem is linearized around the local solution $t_i^{(k)}, a_{ij}^{(k)}, e_{ij}^{(k)}, I_{ij}^{(k)}$ using the difference variables $\tau_i^{(k)}, \alpha_{ij}^{(k)}$.

The eccentricity and inclinations are determined once and for all at the initialization step as explained in Part III : $\begin{cases} e_{ij}^{(k)} = e_{ij}^{(0)} = 0 & \text{(minimum RAAN correction duration)} \\ I_{ij}^{(k)} = I_{ij}^{(0)} & \text{(minimum cost and minimum RAAN correction duration)} \end{cases}$

The iteration process is depicted on Figure 15.

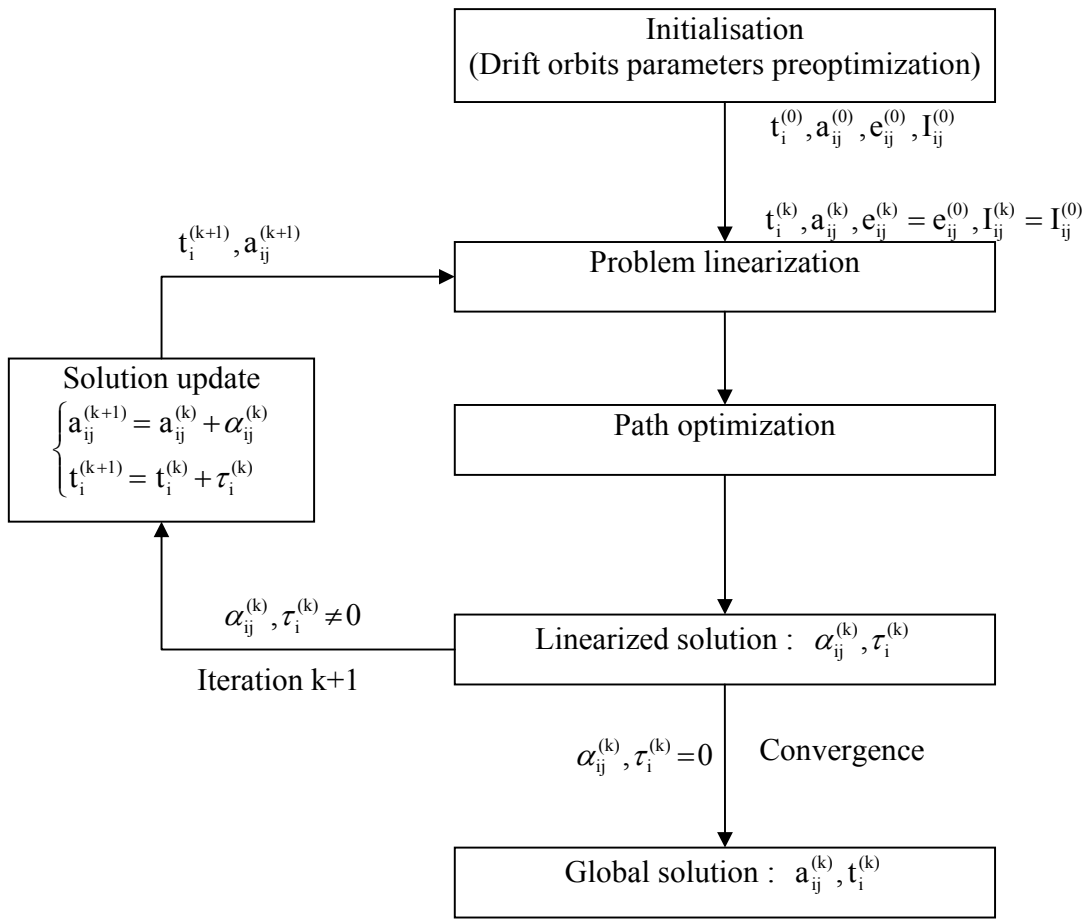

**Figure 15 : Iteration process**

Two successive phases are observed throughout the iterations :

- The first phase starts from the pre-optimized solution with large linearization intervals. During that phase the drift orbits parameters may vary widely from one iteration to the next one. Consequently the optimal path may change. As long as the path changes, it is better to keep the linearization intervals large, or even not to change the initial ones.
- Once the optimal path stabilizes, the second phase consists in converging accurately on the mission duration constraint. Indeed large linearization intervals have led to an over-constrained solution, since the duration linearization function was an overestimate of the true one. This convergence is obtained by reducing progressively the linearization intervals, for example by halving them. Although this was not observed on practical cases, the path may again change during that phase : in that case it is better to restore larger linearization intervals until the stabilization on a new path.

When the convergence on the mission duration is completed, one must check the solution optimality wrt the choices made at the initialization step regarding the side (lower or higher) of the transfers orbit wrt to targeted debris orbit ($a_d$) and the inclination of the transfer orbit. The initial choices were based on a pre-optimization of each transfer with a fixed duration upper bound $T_0$. For different values of $T_0$, the optimal sense for the RAAN correction may be inverted, and these choices must be changed. If the linearized solution exhibits drift altitudes on their bounds, it indicates that the iterations should be resumed with updated assumptions. In practical cases, this was not observed. The debris selected on the initial path are indeed very close in terms of RAAN values, so that the initial sense of RAAN correction remains valid at the end of the iterations.

### D) Transfers re-optimization

Solving the simplified problem yields an optimal paths passing through selected debris. This path has been obtained assuming impulsive manoeuvres to go to and from the drift orbits. In order to get more realistic trajectories, a re-optimization of the manoeuvres and dates can be performed considering a continuous thrust modelling. The path is fixed (debris selection and order) so that the problem is becomes a classical optimal control problem with continuous variables and command laws. The path found with the impulsive approximation can be considered as optimal as long as the Hohmann transfers duration (thrusting manoeuvres and coasts arcs) remains small (less than a few days) wrt the drift durations (typically several weeks). In the case of a very low thrust engine, this approach is no longer valid, since the RAAN corrections must be performed simultaneously with the other orbital parameters corrections. A specific formulation will have to be devised for these difficult problems.

### E) Algorithms

The solving process requires three optimization algorithms respectively for the problem initialization and post-optimization, the linearized problem resolution and the Branch and Bound search.

**Problem initialization**

For every pair of debris, the transfer pre-optimization is a small size (3 variables, 1 constraint) nonlinear problem. A reduced gradient method is used. In order to ensure the solution robustness, several optimizations are repeated for the same problem starting from different initialization values. The best result is retained as initial reference solution for the linearization.

**Branch and Bound**

For a problem with binary variables, the set of all possible combinations is represented as a binary tree[16]. A node of the tree is an instance of the optimization problem where part of the binary variables has been fixed either to 0 or to 1. The root node corresponds to the fully relaxed problem, i.e. with the binary variables considered as real ones. Each node is separate into two children who correspond to fixing a selected binary variable at 0 or 1. These relaxed instances are thus of decreasing dimension when going downwards since more and more binary variables become fixed.

Since the whole tree can potentially be explored, the search strategy must be implemented using as little computer memory as possible[5]. For that purpose only the active nodes (i.e. the nodes still not fully separated in their two children) are stored in a stack. Each node is dynamically allocated as a derived data type, containing the relaxed problem data and a pointer linking the node either to its parent (depth search) or to the next best one in the stack (breadth search). The linked list of active node starts with only one node (root node), is updated by eliminations and/or separations as the search goes on, until it becomes empty when the tree exploration is completed.

The principle of Branch and Bound methods consists in exploring the binary tree downwards from the root node[14,17]. Separating a node consists is solving the associated relaxed problem, where a part of the binary variables are fixed to 0 or 1, while the others are treated as real variables. After the relaxed problem is solved, the following situations may occur :

- The relaxed problem yields a feasible solution wrt the integrity constraints. This solution is compared to the best feasible solution already available from the previous nodes. If better, this new best solution is stored replacing the previous one. There can be no other solution in the branches outgoing from that node. The node is pruned and all its outgoing sub-branches are cut off.

- The relaxed problem is unfeasible because the fixed binary variables are not compliant with the constraints, or the relaxed solution is worse than the reference best solution stored. There can be no better solution in the branches outgoing from that node. The node is pruned and all its outgoing sub-branches are cut off.

- The relaxed problem yields a solution that is neither feasible wrt the integrity constraints nor worse than the reference best solution. The exploration is continued from that node by separating it into two children. Each child is a copy of the parent node with one selected variable (the separation variable) being fixed respectively to 0 and 1.

Having a good reference solution as soon as possible is desirable in order to speed up the resolution by branches cut offs. For that purpose a greedy solution is built before starting the tree exploration by selecting the best arrangement meeting the mission constraint. This solution is stored as initial reference solution.

The critical factors for the efficiency of a Branch and Bound method are :

- The tree exploration strategy ("Branch")
- The node evaluation function ("Bound").

The exploration strategy specifies the way to choose the next node to separate and the separation variable. The depth search strategy consists in separating always the most downwards node in the hope to reach quickly a bottom node of the tree and issue a first reference solution. The breadth search strategy consists in separating the active node with the best evaluation hoping that the best feasible solution lies in a sub-branch of that node (Figure 16). Several possible choices of the separation variable are among others : by numerical order (the most simple), or by constraints (the variable appearing in the maximum number of constraints), or by cost penalty (the variable producing the largest variation in the cost function when changed from 0 to 1). The separation rules (nodes and variables) must be tried case by case in order to assess their practical efficiency on a given problem. The number of nodes examined before issuing the problem solution may vary by large factors depending on these choices and the problem features.

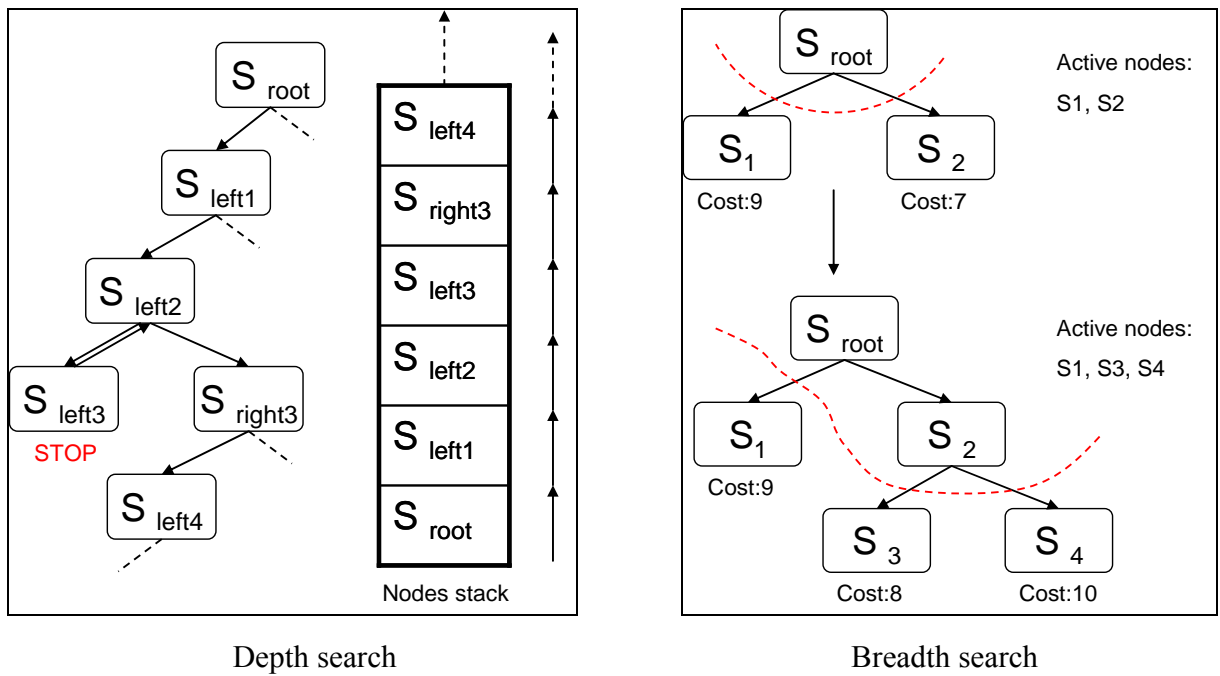

**Figure 16 : Depth and Breath search strategies**

The node evaluation consists in pricing the node, i.e. giving an assessment of the best solution that could be found among all the node children. There is a compromise to make between the time to evaluate the node, the evaluation accuracy and robustness. The goal is to assess as precisely as possible the best solution contained in the node, without consuming an excessive computation time. Solving the node relaxed problem by linear programming is the most robust approach.

**Linear programming**

Three linear programming algorithms are used within the Branch and Bound process to solve the relaxed problems associated to the successive tree nodes :

- A primal simplex to solve the root problem at the top of the tree, when starting from scratch.
- A dual simplex to solve the successive nodes. The parent node solution is used to restore a feasible dual basis, taking into account the additional binary fixed variable chosen to separate the node[16]. This warm-start procedure avoids the risk not to find an initial feasible basis for the current node. It results in robustness and computation time gains.
- An interior point method as backup algorithm. When the simplex fails, either because of numerical rounding errors or degeneracy, the interior point solver is called to restart the node resolution. Similarly to the dual simplex, the parent node solution is retrieved as initialization. This interior point solver is generally more robust, but slower and less accurate than the simplex on medium size problems.

Several techniques of reduction[33] (like variables fixing by constraints elimination) are applied to reduce the size of the linear problem before trying to solve it.

**Selected strategy**

For the SDC mission study, several strategies have been tried in order to select the most efficient one. On the practical run cases the breadth search strategy coupled with the separation on the most constrained variable has given the best performances in terms of number of nodes assessed. The depth and breadth strategies performances are compared for the application case presented in Part V.

### F) Assumptions and simplifications recap

The method presented in the previous sections yields a valid solution (from the optimality point of view) under the following assumptions :

- The mission starts from the first debris selected on the optimal path. The cost and duration required to reach that debris are null, assuming that a launcher has performed the required maneuvers. It is possible to release this assumption, by adding a fictitious starting debris located on the launcher injection orbit.

- The vehicle uses a high thrust engine, so that the maneuvers durations are negligible wrt the coast arcs durations. An impulsive modeling is then representative and the problem becomes of finite dimension.

- The global duration constraint is weak, allowing a drift strategy in order to perform the RAAN corrections at null cost.

- The durations of the debris operations (capture, deorbitation, release) are assumed to be negligible wrt to the other mission phases. The date of arrival and departure from a debris are then assumed to be identical.

- The drift orbits are circular, with their inclinations fixed at the pre-optimization phase. The correctness of the inclination choice is checked a posteriori on the solution. Allowing non null eccentricities could improve the global optimum, at the expenses of increasing the problem dimension.

- The contributions of the mission phases to the global cost and duration are summarized on Table 4.

|  | Cost | Duration |
|---|---|---|
| Transfer manoeuvres | Counted | Neglected |
| Drift phases | Null | Counted |
| Debris operations | Counted | Neglected |

**Table 4 : Cost and duration contributors**

### G) Current status and further work

The solving method has been automated. The initialization step, the problem linearization and the branch and bound resolution are processed sequentially and iterated until convergence. The user needs only to control the validity of each iteration result before running the next one, and if necessary restrict the linearization intervals. Medium size problems like the one presented in Part V can be thus solved in a short time, allowing sensitivities assessments to the mission data.

The enhancements envisioned are the following :

- Allowing non circular drift orbits may reduce the mission cost. This additional degree of freedom leads to an increase of the problem dimension and it will slow down the branch and bound resolution. If a SDC mission for debris on elliptic orbits (like launcher upper stages left on GTO) was envisioned, this enhancement would become necessary. Presently the targeted debris are on LEO and SSO. Before modifying the transfer strategy, it must be checked on a continuous problem (fixed path) if significant performance gains can be expected from elliptic drift orbits.

- Considering non impulsive manoeuvres is possible from the initialization step. The linearized cost and duration function resulting from the transfer pre-optimization will then be more representative of the vehicle thrusting capabilities. The same solving method can then be applied as long as the durations of the transfer manoeuvres remain small wrt the drift durations. If not, a new formulation must be devised taking into account the coupling between the transfer and the drift phases since a part of the RAAN correction is realized during the transfer. The present approach using an intermediate drift orbit is therefore valid only for high or moderate thrust engine.

- Refining the duration function linearization, with two half segments would help reducing the number of iterations, at the expense of larger size problems. The practical run cases have shows that the convergence is typically obtained in less than 10 iterations (6 for the one presented in Part V). This enhancement seems for the moment not useful, but it should be re-examined if the transfer strategy model changes, either to introduce elliptical drift orbits, or to take into account non impulsive manoeuvres.

# V APPLICATION CASE

In order to illustrate the solving method, we consider an application case consisting in selecting 5 debris to deorbit among a list of 11. The problem initial dimensions (Equation 22) are for N=11 :

- 154 binary variables
- 341 real variables
- 1070 constraints

## A) Debris orbits

The 11 debris are on sun-synchronous orbits with regularly spaced inclinations ranging from 98 to 99 deg, and initial RAAN ranging from 160.2 to 235 deg at the date of the mission beginning. The debris orbital parameters are given in Table 5.

|  | Semi-major axis (km) | Eccentricity (-) | Inclination (deg) | Initial RAAN (deg) |
|---|---|---|---|---|
| Debris 1 | 7030.5 | 0.0001 | 98.0 | 221.1 |
| Debris 2 | 7055.3 | 0.0001 | 98.1 | 188.3 |
| Debris 3 | 7080.0 | 0.0001 | 98.2 | 164.4 |
| Debris 4 | 7104.4 | 0.0003 | 98.3 | 235.0 |
| Debris 5 | 7128.5 | 0.0000 | 98.4 | 174.7 |
| Debris 6 | 7152.5 | 0.0001 | 98.5 | 194.1 |
| Debris 7 | 7176.3 | 0.0001 | 98.6 | 149.0 |
| Debris 8 | 7200.0 | 0.0001 | 98.7 | 180.3 |
| Debris 9 | 7223.2 | 0.0002 | 98.8 | 200.6 |
| Debris 10 | 7246.4 | 0.0001 | 98.9 | 191.0 |
| Debris 11 | 7269.3 | 0.0003 | 99.0 | 160.2 |

**Table 5 : List of 11 candidate debris**

The altitude of the drift orbits is bounded in the range 400 km – 1200 km.

### B) Initialization

Every transfer from any debris to any other is pre-optimized for a fixed duration of 2 months. Once all the optimizations are completed, the unfeasible transfers are discarded from the problem by setting their selection variables $s_{ij}$ to 0. The remaining feasible transfers are linearized around their solution in order to initialize the iteration process for the branch and bound. This pre-processing allows a reduction of the problem dimension from 500 to 200 variables and from 1000 to 500 constraints.

At this initialization step, with fixed transfers duration set to 2 months, the optimal path would be : $5 \rightarrow 8 \rightarrow 2 \rightarrow 10 \rightarrow 6$, for a total $\Delta V$ of 711 m/s and a total duration of 8 months (2 months per transfer). The intermediate orbits on this initial path are presented in Table 6.

It can be observed at this initialization steps that the mission cost is driven by the RAAN differences rather than the inclinations or the altitudes. The best strategy consists in selecting the debris with the minimum RAAN differences in order to have drift orbits as close as possible to the debris orbits. The drift orbits are lower than the debris orbits, with higher precession rates, so that the vehicle catches the debris forwards. The debris are therefore ordered by increasing RAAN values.

| Initial path | Semi-major axis (km) | Inclination (deg) | Cost $\Delta V$ (m/s) | Duration $\Delta T$ (days) |
|---|---|---|---|---|
| Debris 5 | 7128.5 | 98.4 | ≈0 | ≈0 |
| Drift 5 → 8 | 7019.6 | 98.7 | 173.3 | 61 |
| Debris 8 | 7200.0 | 98.7 | ≈0 | ≈0 |
| Drift 8 → 2 | 6947.9 | 98.7 | 246.9 | 61 |
| Debris 2 | 7055.3 | 98.1 | ≈0 | ≈0 |
| Drift 2 → 10 | 7140.9 | 98.9 | 176.5 | 61 |
| Debris 10 | 7246.4 | 98.9 | ≈0 | ≈0 |
| Drift 10 → 6 | 7125.5 | 98.9 | 114.1 | 61 |
| Debris 6 | 7152.5 | 98.5 | ≈0 | ≈0 |
| Total | | | 710.8 | 244 |

**Table 6 : Initial path**

### C) Iterations

The initial linearized problem is solved by a branch and bound method starting from the pre-optimized path as feasible initialization. The depth and breadth search strategies are tried at this first iteration in order to determine the most efficient one for the further iterations.

The numbers of nodes assessed are respectively 135 for the depth search and 99 for the breadth search. The difference may be explained by the fact that an already good reference solution is provided by the initial path. The depth search whose goal is to find a feasible solution as soon as possible, while neglecting nodes elimination (opposite to the breadth search) is therefore less efficient. The solutions issued from the successive iterations are presented in Table 7.

| Iteration | Number of nodes | Path | Cost $\Delta V$ (m/s) | Duration $\Delta T$ (days) |
|---|---|---|---|---|
| 0 | 99 | $5 \to 8 \to 2 \to 10 \to 6$ | 710.8 | 244.0 |
| 1 | 53 | $5 \to 8 \to 2 \to 10 \to 6$ | 652.3 | 269.3 |
| 2 | 41 | $5 \to 8 \to 2 \to 10 \to 6$ | 594.8 | 299.1 |
| 3 | 33 | $5 \to 8 \to 2 \to 10 \to 6$ | 540.7 | 335.7 |
| 4 | 41 | $5 \to 8 \to 2 \to 6 \to 10$ | 508.0 | 363.3 |
| 5 | 13 | $5 \to 8 \to 2 \to 6 \to 10$ | 502.8 | 364.1 |
| 6 |  | $5 \to 8 \to 2 \to 6 \to 10$ | 500.7 | 366.0 |

**Table 7 : Iterations**

The optimal path changes at the iteration 4 with a permutation of the two last debris, and becomes 5 → 8 → 2 → 6 → 10. This optimal path requires a total ΔV of 500 m/s and a total duration of 12 months (maximal duration allowed).

The iterations require less and less nodes evaluations, while the solution approaches the optimum, because the initial reference solution is better and allows more efficient cut offs. The saturation of the duration constraint was expected since taking as long time as possible for the RAAN correction allow the drift orbits to be close to the debris orbits, and thus minimizes the manoeuvres cost.

### D) Optimal solution

The intermediate orbits on the optimal path are presented in Table 8.

It is observed at the optimal debris order is no longer by increasing RAAN values. It proves more favourable to change the sense of the RAAN correction for the last transfer, in order to reduce the inclination change manoeuvres.

Taking into account the deorbitation manoeuvre (if the vehicle has to deorbit the debris) does not change the optimal path. Indeed the debris being on very close circular orbits, the deorbitation cost is quasi identical for all of them, and it does not influence the debris choice. In order to ensure a natural fall out, the deorbitation manoeuvre must lower the perigee inside the atmosphere. The order of magnitude of this manoeuvre is 200 m/s per debris, so that the total mission cost increase amounts to 1000 m/s.

| Optimal path | Semi-major axis (km) | Inclination (deg) | Cost ΔV (m/s) | Duration ΔT (days) |
|---|---|---|---|---|
| Debris 5 | 7128.5 | 98.4 | 0 | 0 |
| Drift 5 → 8 | 7090.6 | 98.7 | 107.9 | 103.0 |
| Debris 8 | 7200.0 | 98.7 | 0 | 0 |
| Drift 8 → 2 | 7042.6 | 98.7 | 165.2 | 100.8 |
| Debris 2 | 7055.3 | 98.1 | 0 | 0 |
| Drift 2 → 6 | 7028.2 | 98.5 | 126.4 | 92.8 |
| Debris 6 | 7152.5 | 98.5 | 0 | 0 |
| Drift 6 → 10 | 7247.7 | 98.5 | 101.2 | 69.4 |
| Debris 10 | 7246.4 | 98.9 | 0 | 0 |
| Total | | | 500.7 | 366 |

**Table 8 : Optimal path**

The vehicle trajectory (drift orbits), the RAAN evolution (difference between the vehicle RAAN and the debris RAAN), and the mission cost (before and after optimization) are plotted respectively on Figure 17, Figure 18 and Figure 19.

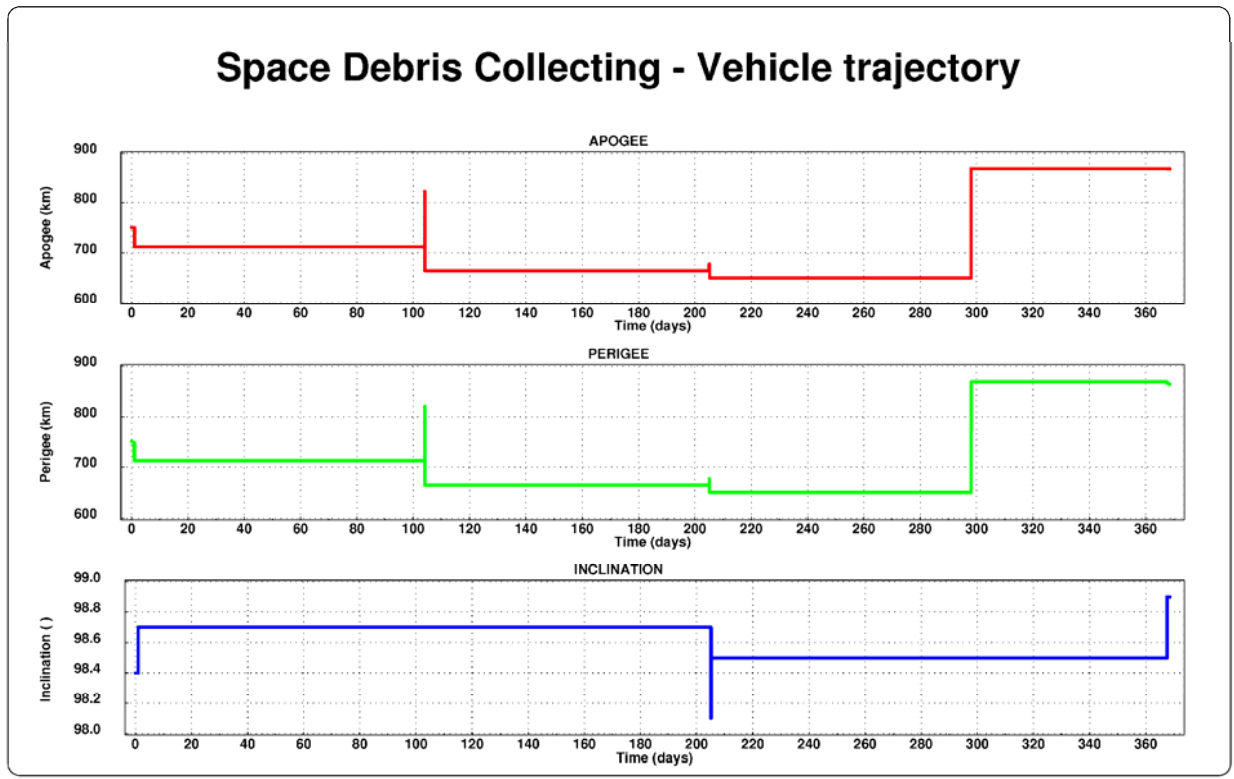

**Figure 17 : Optimal path orbits**

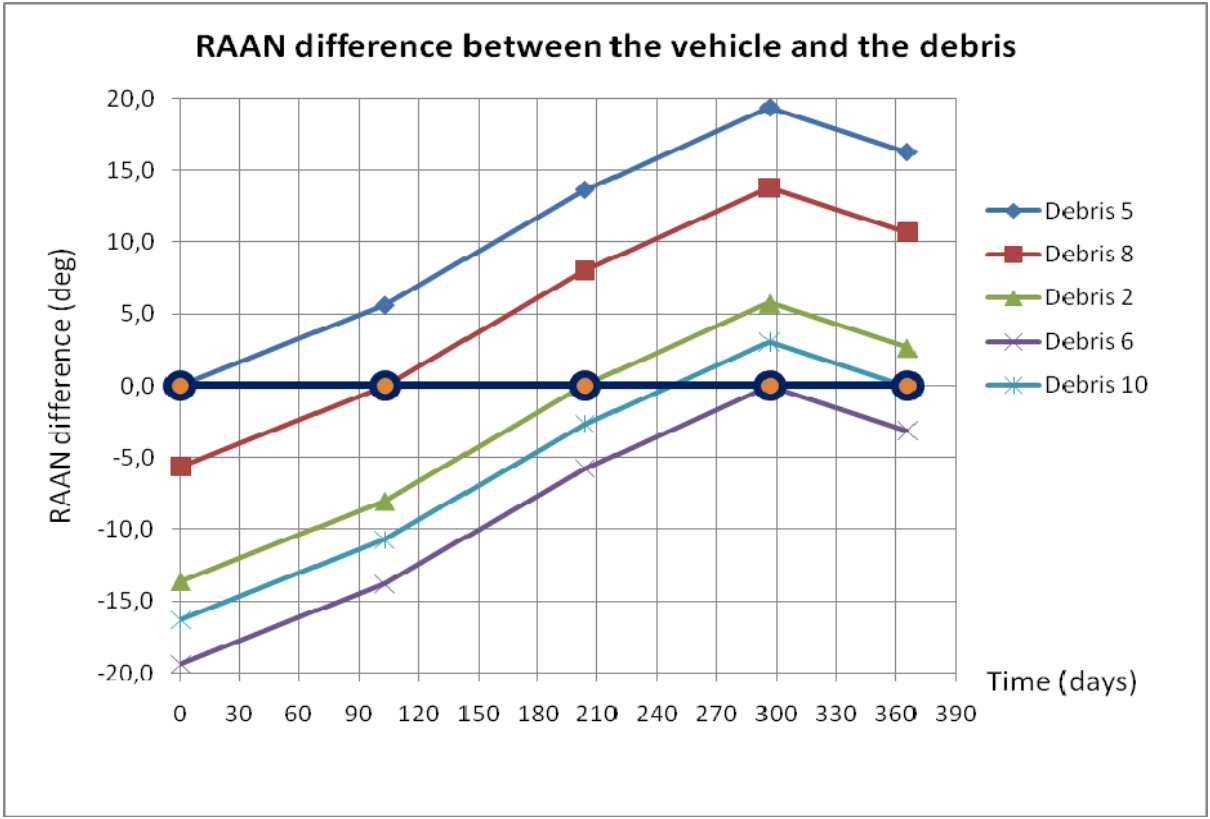

**Figure 18 : RAAN correction**

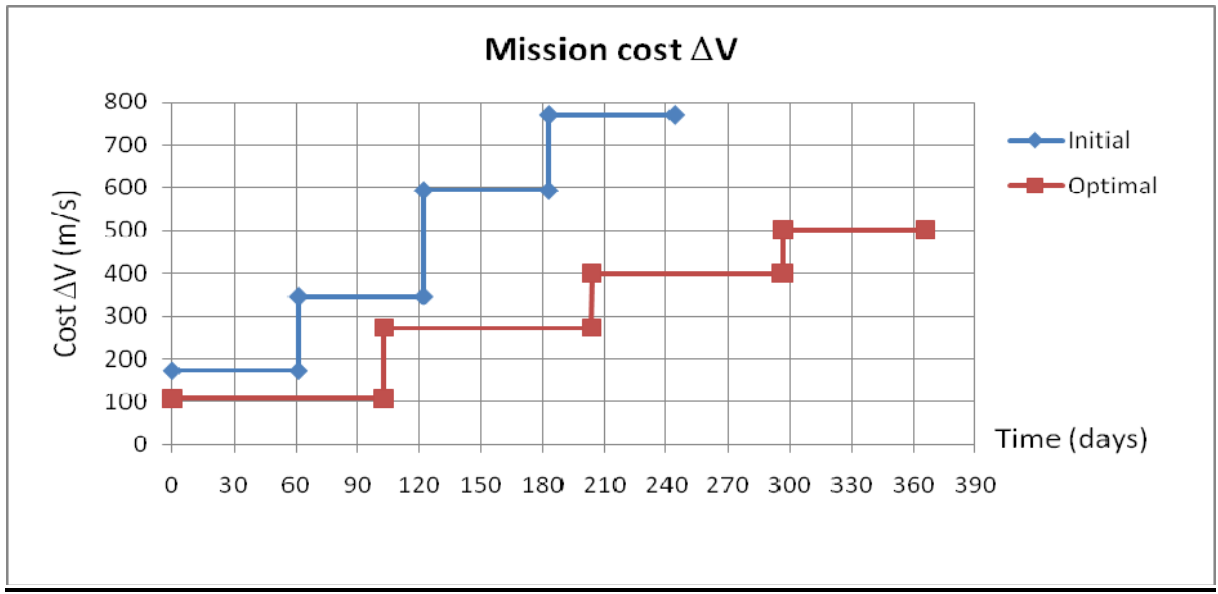

**Figure 19 : Mission cost before and after optimization**

# VI  CONCLUSION

This paper proposes a method to optimize simultaneously the debris selection and the trajectories between them in view of a Space Debris Collecting mission. A generic transfer strategy with impulsive maneuvers is defined so that the problem becomes of finite dimension. The problem is then linearized around an initial reference solution in order to apply a Branch and Bound algorithm. The process is iterated until the solution stabilizes on the optimal path. The method is applicable whatever the numbers of debris (candidate and to deorbit) and whatever the mission duration. The main practical limitation comes from the problem size and the associated computation time that grow exponentially with the number of candidate debris. The initialization procedure includes a filtering of the unfeasible or too expensive solutions, allowing thus a subsequent size reduction and a time-efficient numerical resolution.

An application case consisting in selecting 5 SSO debris among a list on 11 candidates is presented. The method proves reliable and an optimal path is issued in a few iterations. The optimal path can then be used as a basis for a more detailed mission analysis, taking into account the vehicle features (thrust level) and the operational constraints (rendezvous and deorbitation operations). Sensibilities to the mission main constraints (duration, altitudes bounds) can also be easily issued in order to support design trade-offs for a future SDC vehicle.

The main enhancements envisioned are to allow non circular drift orbits and to account for non impulsive manoeuvres in the cost and durations assessment. The two interests in allowing elliptical drift orbits are : to reduce the SDC mission cost for LEO and SSO debris, and to apply the method to GTO debris which are on highly elliptical orbits. Considering continuous instead of impulsive thrusting is more representative of the vehicle capabilities when assessing the mission cost. This is possible without changing the solving method as long as the manoeuvres durations remain small wrt the drift phase durations. For very low-thrust engine, the transfer and the drift phases become highly coupled, since a part of the RAAN correction is realized during the transfer. A specific transfer strategy has to be devised for this category of vehicles.

# REFERENCES

1. "Position paper on space debris mitigation". (2005). International Academy of Astronautics.
2. B. Bastida Virgili, H. Krag. (2009). "Strategies for active removal in LEO".
3. Billionnet, A. (2007). *"Optimisation discrète"*. Dunod.
4. Chao, C. (2005). *Applied orbit perturbation and maintenance.* The Aerospace Press.
5. Chapman, S. J. (2004). *"Fortran 90/95 for Scientists and Engineers - Second Edition"*. McGraw Hill.
6. Chobotov, V. (2002). *"Orbital mechanics - Third Edition"*. AIAA Education Series.
7. G.L. Nemhauser, A.H.G Rinnooy Kan, M.J. Todd. (1989). *"Optimization ", Handbooks in operations research and management science, Volume 1.* Elsevier.
8. H.J. Oberle, K. Taubert. (1997). "Existence and multiple solutions of the minimum-fuel orbit transfer problem". *J. Optim Theory Appl. 95* , 243-262.
9. J. Dréo, A. Pétrowski, P. Siarry, E. Taillard. (2003). *"Métaheuristiques pour l'optimisation difficile"*. Eyrolles.
10. J. Gergaud, T. Haberkorn. (2007). "Orbital transfer : some links between the low-thrust and the impulse cases". *Acta Astronautica, 60, no. 6-9* , 649-657.
11. J. Murakami, S. Hokamoto. (2010). "Approach for optimal multi-rendezvous trajectory design for active debris removal". *61th International Astronautical Congress*, (pp. IAC-10.C1.5.8). Prague, CZ.
12. J.J. Schneider, S. Kirkpatrick. (2006). *"Stochastic optimization"*. Springer.
13. Klinkrad, K. (s.d.). "Space debris models and risks analysis". *ISBN: 3-540-25448-X* .
14. Leitmann, G. (1962). *"Optimization techniques", Mathematics in science and engineering, Volume 5.*
15. Liou, J. (2008). "An assessment of the current LEO debris environment and what needs to be done to preserve it for future generations".
16. Minoux, M. (2008). *"Programmation mathématique - Théorie et algorithms – 2ème Edition"*. Editions Tec&Doc – Lavoisier.
17. P. Lacomme, C. Prins, M. Sevaux. (2003). *"Algorithmes de graphes– 2ème Edition "*. Eyrolles.
18. R. Walker, C.E. Martin, P.H. Stokes, J.E. Wilkinson. (2000). "Studies of space debris mitigation options using the debris environment long term analysis (DELTA) model". *51st International Astronautical Congress, Brazil*, (pp. IAA-00_IAA.6.6.07). Rio de Janeiro.
19. Trélat, E. (2005). *"Contrôle optimal - Théorie et applications"*. Vuibert.

# LIST OF TABLES

# LIST OF FIGURES

# ACRONYMS

LEO :        Low Earth Orbit

SSO :        Sun-Synchronous Orbit

GTO :        Geostationary Earth Orbit

SDC :       Space Debris Collecting
TSP :       Travelling Salesman Problem
PMP :       Pontryaguin Maximum Principle
NLP :       Non Linear Programming
LP :        Linear Programming
BB :        Branch and Bound
AN :        Ascending Node
RAAN :      Right Ascension of the Ascending Node

## VARIABLES NOMENCLATURE

| Vehicle | | | |
|---|---|---|---|
| $X(t)$ | State vector (position + velocity + mass) | Real | 7 |
| $m(t)$ | Mass | Real | 1 |
| $Y(t)$ | State vector (position + velocity) | Real | 6 |
| $U(t)$ | Command vector | Real | 3 |

| Mission | | | |
|---|---|---|---|
| N | Number of candidate debris | Integer | 1 |
| n | Number of debris to deorbit | Integer | 1 |
| $T_{deorb}$ | Deorbitation operations duration | Real | 1 |
| $T_{min}$ | Transfer minimum duration | Real | 1 |
| $T_{max}$ | Mission maximum duration | Real | 1 |
| C | Mission cost | Real | 1 |
| T | Mission duration | Real | 1 |
| $C_L$ | Mission linearized cost | Real | 1 |
| $T_L$ | Mission linearized duration | Real | 1 |

| | Orbits | | |
|---|---|---|---|
| $R_T$ | Earth equatorial radius (= 6378137 m) | Real | 1 |
| $\mu$ | Earth gravitational constant (= 3.986 $10^{14}$ m³/s²) | Real | 1 |
| $J_2$ | First zonal coefficient (= 1.086 $10^{-3}$) | Real | 1 |
| $C_{J2}$ | $C_{J2} = \frac{3}{2} J_2 \sqrt{\mu} \, R_T^2 = 1.318 \, 10^{18}$ | Real | 1 |
| a | Semi major axis | Real | 1 |
| e | Eccentricity | Real | 1 |
| I | Inclination | Real | 1 |
| $\Omega$ | Right ascension of the ascending node | Real | 1 |
| $\omega$ | Argument of the perigee | Real | 1 |
| $\nu$ | True anomaly | Real | 1 |
| L | Longitude from the ascending node | Real | 1 |
| n | Mean motion | Real | 1 |
| $\dot{\Omega}$ | RAAN precession rate | Real | 1 |
| H | Altitude | Real | 1 |
| V | Velocity | Real | 1 |
| $\Delta V$ | Impulsive velocity | Real | 1 |
| $\Delta V_P$ | Perigee impulsive velocity (Hohmann transfer) | Real | 1 |
| $\Delta V_A$ | Apogee impulsive velocity (Hohmann transfer) | Real | 1 |

| Debris k, k=1 to N | | | |
|---|---|---|---|
| $s_k$ | Debris selection | Binary | 1 |
| $x_k$ | Number of arriving edges | Binary | 1 |
| $y_k$ | Number of departing edges | Binary | 1 |
| $z_k$ | Product of $x_k$ and $y_k$ | Binary | 1 |
| $t_k^a$ | Arrival date | Real | 1 |
| $t_k^d$ | Departure date | Real | 1 |
| $C_k$ | Operations cost | Real | 1 |
| $T_k$ | Operations duration | Real | 1 |
| $a_k$ | Semi major axis | Real | 1 |
| $Y_k(t)$ | State vector (position + velocity) | Real | 6 |
| $e_k$ | Eccentricity | Real | 1 |
| $I_k$ | Inclination | Real | 1 |
| $\Omega_k$ | Right ascension of the ascending node | Real | 1 |
| $\dot{\Omega}_k$ | RAAN precession rate | Real | 1 |

| Transfer i-j, i,j=1 to N, i≠j | | | |
|---|---|---|---|
| $s_{ij}$ | Transfer selection | Binary | 1 |
| $C_{ij}$ | Transfer cost | Real | 1 |
| $T_{ij}$ | Transfer duration | Real | 1 |
| $t_i$ | Departure date | Real | 1 |
| $t_j$ | Arrival date | Real | 1 |
| $C_{Lij}$ | Transfer linearized cost | Real | 1 |
| $T_{Lij}$ | Transfer linearized duration | Real | 1 |
| $a_{ij}$ | Drift orbit semi major axis | Real | 1 |
| $e_{ij}$ | Drift orbit eccentricity | Real | 1 |
| $I_{ij}$ | Drift orbit inclination | Real | 1 |
| $\dot{\Omega}_{ij}$ | Drift orbit RAAN precession rate | Real | 1 |
| $\Delta V_{ij}$ | Transfer impulsive velocity | Real | 1 |
| $\alpha_{ij}$ | Semi major axis difference wrt reference | Real | 1 |
| $\tau_i$ | Departure date difference wrt reference | Real | 1 |
| $p_{ij}$ | Product of $s_{ij}$ and $a_{ij}$ | Real | 1 |
| $q_{ij}$ | Product of $s_{ij}$ and $t_i$ | Real | 1 |
| $a_{min}, a_{min}$ | Bounds on semi major axis | Real | 1 |
| $\alpha_{min}, \alpha_{min}$ | Bounds on semi major axis difference | Real | 1 |
| $\tau_{min}, \tau_{min}$ | Bounds on departure date difference | Real | 1 |